\documentclass[12pt]{article}
\usepackage{graphicx}
\def\beq{\begin{equation}}
\def\eeq{\end{equation}}
\def\bea{\begin{eqnarray}}
\def\eea{\end{eqnarray}}
\def\nn{\nonumber}

\def\roughly#1{\mathrel{\raise.3ex\hbox
{$#1$\kern-.75em\lower1ex\hbox{$\sim$}}}}

\def\sla#1{\raise.15ex\hbox{$/$}\kern-.57em #1}
\def\bra#1{\left\langle #1\right|}
\def\ket#1{\left| #1\right\rangle}

\def\bd{B_d^0}

\def\pewcp{P_{EW}^{\prime C}}
\def\pewp{P'_{EW}}

\def\btos{{\bar b} \to {\bar s}}
\def\order{\lower 1.8ex \hbox{\LARGE\~{}}}
\def\btokpi{B \to K \pi}
\def\btokpipi{B \to K \pi \pi}

\makeatletter
\newcommand{\contraction}[5][1ex]{%
 \mathchoice
  {\contraction@\displaystyle{#2}{#3}{#4}{#5}{#1}}%
  {\contraction@\textstyle{#2}{#3}{#4}{#5}{#1}}%
  {\contraction@\scriptstyle{#2}{#3}{#4}{#5}{#1}}%
  {\contraction@\scriptscriptstyle{#2}{#3}{#4}{#5}{#1}}}%
\newcommand{\contraction@}[6]{%
 \setbox0=\hbox{$#1#2$}%
 \setbox2=\hbox{$#1#3$}%
 \setbox4=\hbox{$#1#4$}%
 \setbox6=\hbox{$#1#5$}%
 \dimen0=\wd2%
 \advance\dimen0 by \wd6%
 \divide\dimen0 by 2%
 \advance\dimen0 by \wd4%
 \vbox{%
  \hbox to 0pt{%
   \kern \wd0%
   \kern 0.4\wd2
   \contraction@@{\dimen0}{#6}%
   \hss}%
  \vskip 0.8ex
  \vskip\ht2}}

\newcommand{\contraction@@}[3][0.06em]{%
 \hbox{%
  \vrule width #1 height 0pt depth #3%
  \vrule width #2 height 0pt depth #1%
  \vrule width #1 height 0pt depth #3%
  \relax}}
\makeatother
\begin{document}

\begin{flushright}
UdeM-GPP-TH-10-195 \\
\end{flushright}

\begin{center}
\bigskip
{\Large \bf \boldmath Measuring $\gamma$ in $\btokpipi$ Decays} \\
\bigskip
\bigskip
{\large 
Maxime Imbeault $^{a,}$\footnote{imbeault.maxime@gmail.com},
Nicolas Rey-Le Lorier $^{b,}$\footnote{nicolas.rey-le.lorier@umontreal.ca},
and David London $^{b,}$\footnote{london@lps.umontreal.ca}
}
\end{center}

\begin{flushleft}
~~~~~~~~~~~$a$: {\it D\'epartement de physique, C\'egep de Baie-Comeau,}\\
~~~~~~~~~~~~~~~{\it 537 boulevard Blanche, Baie-Comeau, QC, Canada G5C 2B2}\\
~~~~~~~~~~~$b$: {\it Physique des Particules, Universit\'e
de Montr\'eal,}\\
~~~~~~~~~~~~~~~{\it C.P. 6128, succ. centre-ville, Montr\'eal, QC,
Canada H3C 3J7}\\
\end{flushleft}

\begin{center}
\bigskip (\today)
\vskip0.5cm {\Large Abstract\\} \vskip3truemm
\parbox[t]{\textwidth}{We re-examine the question of measuring the
  weak phase $\gamma$ in $\btokpipi$ decays. To this end, we express
  all $\btokpipi$ amplitudes in terms of diagrams. We show that, as in
  $\btokpi$, there exist relations between certain tree and
  electroweak-penguin diagrams. The imposition of these relations
  allows the extraction of $\gamma$ from measurements of the
  $\btokpipi$ observables. We estimate the theoretical error in this
  method to be $O(5\%)$.}
\end{center}

\thispagestyle{empty}
\newpage
\setcounter{page}{1}
\baselineskip=14pt

\section{Introduction}

In the standard model, CP violation is due to a phase in the
Cabibbo-Kobayashi-Maskawa (CKM) quark mixing matrix. The CKM phase
information is conventionally parametrized in terms of the unitarity
triangle, in which the interior (CP-violating) angles are known as
$\alpha$, $\beta$ and $\gamma$ \cite{pdg}. In this paper, we discuss a
method for measuring $\gamma$ in $\btokpipi$ decays.  In order to
put this discussion into context, we begin with a review of weak
phases in $\btokpi$.

At the end of the 1980's, it was thought that $\btokpi$ receives
contributions only from tree-type diagrams (proportional to
$e^{i\gamma}$) and penguin diagrams (no weak phase). The appearance of
two contributions with different weak phases meant that it was not
possible to obtain clean weak-phase information from the measurement
of the indirect CP asymmetry. In 1991, Nir and Quinn (NQ) \cite{NQ}
showed that one can use an isospin analysis to eliminate the ``penguin
pollution,'' so that one could indeed obtain $\gamma$ from $\btokpi$
decays. However, several years later it was noted that, in fact, these
decays receive significant electroweak-penguin (EWP) contributions
\cite{DH}, and that their appearance makes the NQ analysis fail.
Several years after that, it was shown that, under flavor SU(3)
symmetry, the EWP diagrams are proportional to the tree diagrams
(apart from their weak phases) \cite{NR,GPY}. Finally, in 2004, all
this information was put together, and it was found that it is
possible to modify the NQ analysis using the EWP-tree relations, and
cleanly extract $\gamma$ from $\btokpi$ \cite{Kpisol}.

Weak phases in $\btokpipi$ follow a similar story (up to a
point). (Note: assuming isospin symmetry, the wavefunction in $B \to
K\pi\pi$ decays must be symmetrized with respect to the exchange of
the final-state pions. Depending on their relative angular momentum,
the $\pi\pi$ isospin state must be symmetric or antisymmetric.) In
1991, Lipkin, Nir, Quinn and Snyder (LNQS) performed an isospin
analysis of $K\pi\pi$, and obtained the relations among the amplitudes
for the various $\btokpipi$ decays, for both the symmetric and
antisymmetric cases \cite{LNQS}. Assuming the experimental separation
of these cases, they noted that the relations permit one to extract
clean weak-phase information from $\btokpipi$ decays. However, their
analysis was based in part on that of Nir and Quinn, i.e.\ EWP
contributions were neglected. Once these are included, the LNQS method
fails. In 2003, Deshpande, Sinha and Sinha (DSS) attempted to revive
the LNQS analysis for the case of symmetric $\pi\pi$ isospin states
\cite{DSS}. They included EWPs in a schematic way, and assumed that
these can be related to the tree diagrams, as in
Refs.~\cite{NR,GPY}. Within their assumptions, they argued that it is
possible to extract $\gamma$ from $B \to K\pi\pi$. However, it was
subsequently noted that the assumed EWP-tree relation in $K\pi\pi$
does not hold \cite{Grocomment}, so that we are back to the situation
of being unable to obtain weak-phase information from $\btokpipi$.
This is how things stand presently.

In light of this, in this paper we re-examine the question of whether
it is possible to measure $\gamma$ in $\btokpipi$ decays.  To this
end, we express the $\btokpipi$ amplitudes in terms of diagrams, and
note that the number of unknown theoretical parameters does indeed
exceed the number of observables. Thus, one cannot extract weak phases
without additional information.

This input comes from EWP-tree relations. It is true that the relation
assumed by DSS does not hold. However, we show that there are other
relations between certain EWP and tree diagrams. If these are taken
into account, this reduces the number of unknown theoretical
parameters, so that the extraction of $\gamma$ is possible.
Experimentally, it is not easy, but it is fairly clean theoretically.

In Sec.~2, we introduce the diagrams and show how to express the
$\btokpipi$ amplitudes in terms of these. EWP-tree relations are
discussed in Sec.~3. The contractions formalism is used to derive
such relations for $\btokpipi$ decays. In Sec.~4, we show how the
EWP-tree relations permit the measurement of $\gamma$ in $\btokpipi$
decays. We conclude in Sec.~5.

\section{\boldmath $\btokpipi$ Amplitudes}
\label{amps}

There are six processes in $\btokpipi$ decays: $B^+ \to
K^+\pi^+\pi^-$, $B^+ \to K^+\pi^0\pi^0$, $B^+ \to K^0\pi^+\pi^0$, $\bd
\to K^+\pi^-\pi^0$, $\bd \to K^0\pi^+\pi^-$, $\bd \to K^0\pi^0\pi^0$.
For the moment, we assume only isospin symmetry, as in
Refs.~\cite{LNQS,DSS}. In all of these decays, the overall
wavefunction of the final $\pi\pi$ pair must be symmetrized with
respect to the exchange of these two particles. If the relative
$\pi\pi$ angular momentum is even (odd), the isospin state must be
symmetric (antisymmetric). We refer to these two cases as
$I_{\pi\pi}^{sym}$ and $I_{\pi\pi}^{anti}$.

In Ref.~\cite{BPPP} it was shown that $I_{\pi\pi}^{sym}$ and
$I_{\pi\pi}^{anti}$ can be determined experimentally. We briefly
summarize the argument. Consider, for example, $\bd \to K^0\pi^+\pi^-$
(other decays are treated similarly). The events in the Dalitz plot
can be described by the following two variables:
\bea
s_+ &=& m^2_{K^0\pi^+} = \left( p_{K^0} + p_{\pi^+} \right)^2 ~, \nn\\
s_- &=& m^2_{K^0\pi^-} = \left( p_{K^0} + p_{\pi^-} \right)^2 ~.
\label{s+s-defs}
\eea
Now, a Dalitz-plot analysis permits the extraction of the decay
amplitude, ${\cal M}(s_+,s_-)$, including both resonant and
non-resonant contributions. The key point is that, under the exchange
of the two pions, we have $p_{\pi^+} \leftrightarrow p_{\pi^-}$,
i.e.\ $s_+ \leftrightarrow s_-$. Thus, the symmetric and antisymmetric
amplitudes are simply $\frac{1}{\sqrt{2}}[{\cal M}(s_+,s_-) \pm {\cal
    M}(s_-,s_+)]$.

In fact, the full amplitude cannot be obtained -- its global phase is
undetermined. Thus, it is really $|{\cal M}|$ which is extracted.
Similarly, one can obtain $|{\overline{\cal M}}|$ from the
CP-conjugate decay. Therefore, for each decay one measures the
momentum-dependent branching ratio ($\propto |{\cal M}|^2 +
|{\overline{\cal M}}|^2$) and the momentum-dependent direct CP
asymmetry ($\propto |{\cal M}|^2 - |{\overline{\cal M}}|^2$). In
addition, for $K^0\pi^+\pi^-$ (where the $K^0$ is seen as $K_S$), the
momentum-dependent indirect CP asymmetry\footnote{The indirect CP
  asymmetry depends on the CP of the final state, and a-priori
  $K^0\pi^+\pi^-$ is a mixture of CP $+$ and CP $-$. However, the
  separation of symmetric and antisymmetric $\pi\pi$ states also fixes
  the final-state CP: $K^0(\pi\pi)_{sym}$ and $K^0(\pi\pi)_{anti}$
  have CP $+$ and $-$, respectively.} can be measured, and gives
${\cal M}^* {\overline{\cal M}}$ for this decay.

In the $I_{\pi\pi}^{sym}$ scenario, there are several relations among
the amplitudes, including $A(B^+ \to K^0\pi^+\pi^0)_{sym} = -A(\bd \to
K^+\pi^-\pi^0)_{sym}$ \cite{LNQS}. This implies that there are only
five independent decays. For $I_{\pi\pi}^{anti}$, there are only four
processes: $B^+ \to K^+\pi^+\pi^-$, $B^+ \to K^0\pi^+\pi^0$, $\bd \to
K^+\pi^-\pi^0$, $\bd \to K^0\pi^+\pi^-$ (one cannot antisymmetrize a
$\pi^0\pi^0$ state).

Now, the goal here is to extract the weak phase $\gamma$ from
measurements of $\btokpipi$ decays. This can be done if the number of
unknown theoretical parameters in the amplitudes is less than or equal
to the number of observables. In the $I_{\pi\pi}^{sym}$ case, there
are 11 observables: the momentum-dependent branching ratios and direct
CP asymmetries of $B^+ \to K^+\pi^+\pi^-$, $B^+ \to K^+\pi^0\pi^0$,
$\bd \to K^+\pi^-\pi^0$, $\bd \to K^0\pi^+\pi^-$, $\bd \to
K^0\pi^0\pi^0$, and the momentum-dependent indirect CP asymmetry of
$\bd \to K^0\pi^+\pi^-$ (the indirect CP asymmetry of $\bd \to
K^0\pi^0\pi^0$ will essentially be impossible to measure). For
$I_{\pi\pi}^{anti}$, there are 9 observables: the momentum-dependent
branching ratios and direct CP asymmetries of $B^+ \to K^+\pi^+\pi^-$,
$B^+ \to K^0\pi^+\pi^0$, $\bd \to K^+\pi^-\pi^0$, $\bd \to
K^0\pi^+\pi^-$, and the momentum-dependent indirect CP asymmetry of
$\bd \to K^0\pi^+\pi^-$. We therefore conclude that the
$I_{\pi\pi}^{sym}$ scenario is the more promising for extracting
$\gamma$, and we concentrate on it exclusively below.

\begin{figure}
	\centering
		\includegraphics[height=3.98cm]{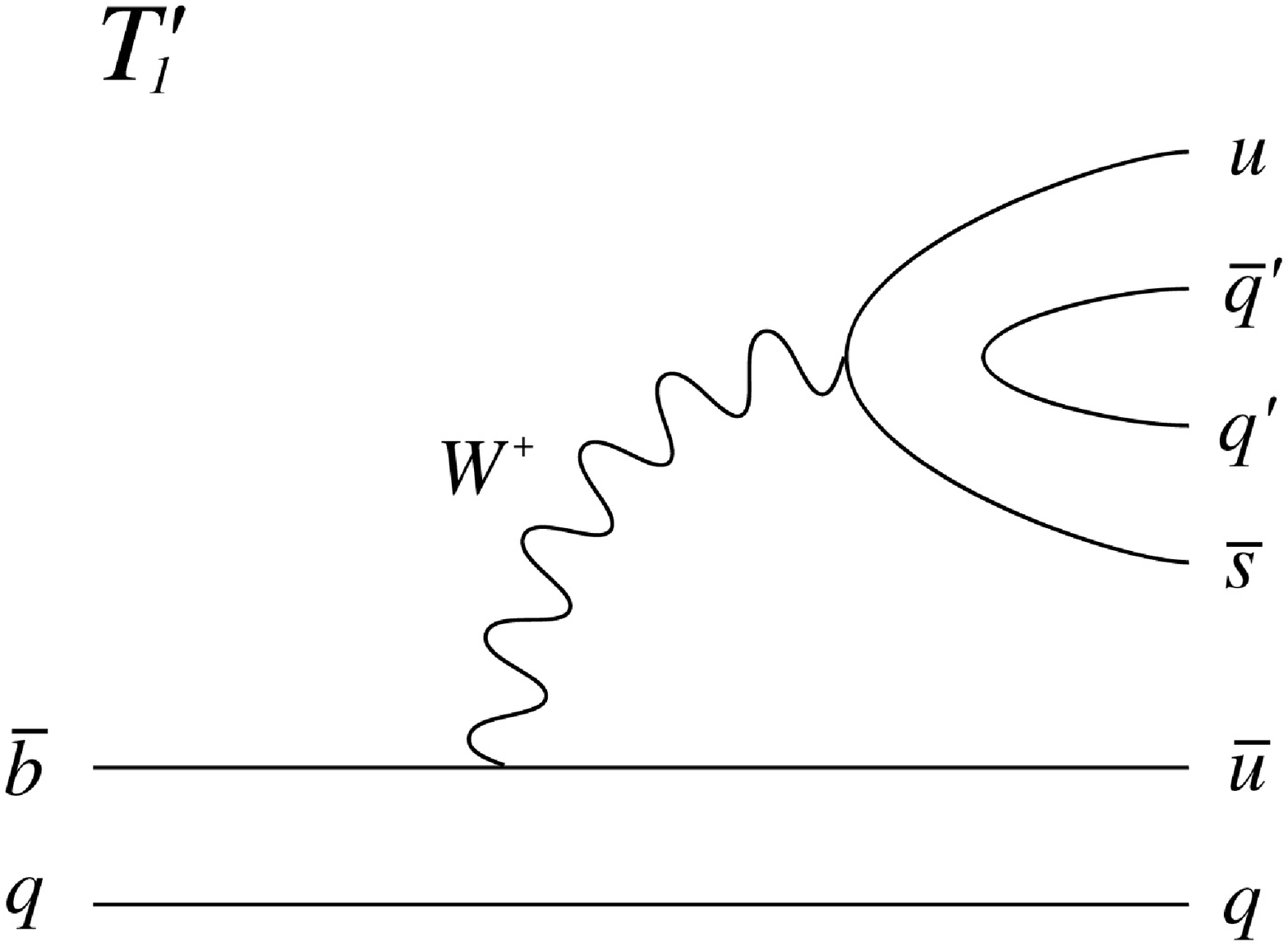}
		\includegraphics[height=3.98cm]{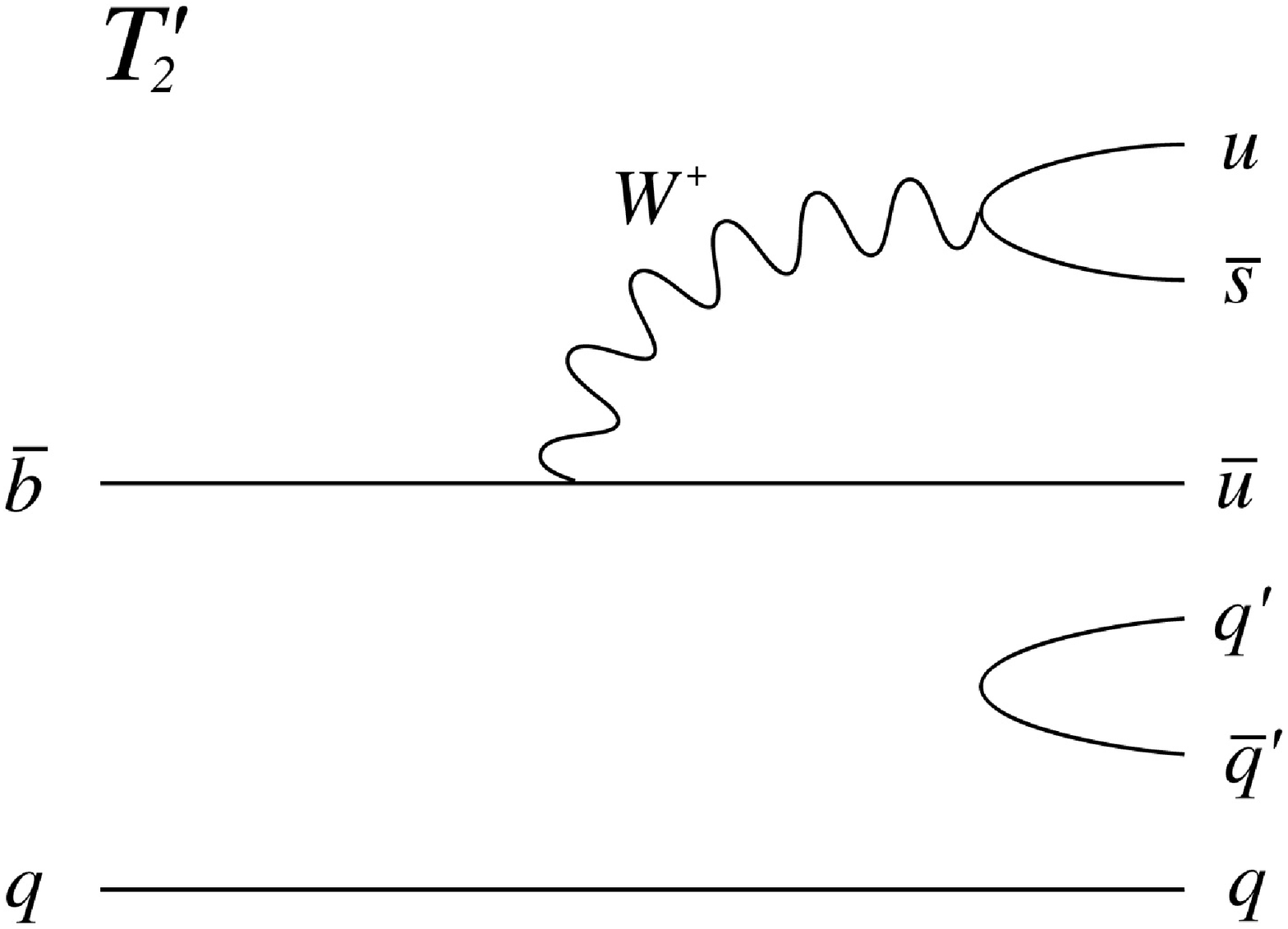}
	\centering
		\includegraphics[height=3.98cm]{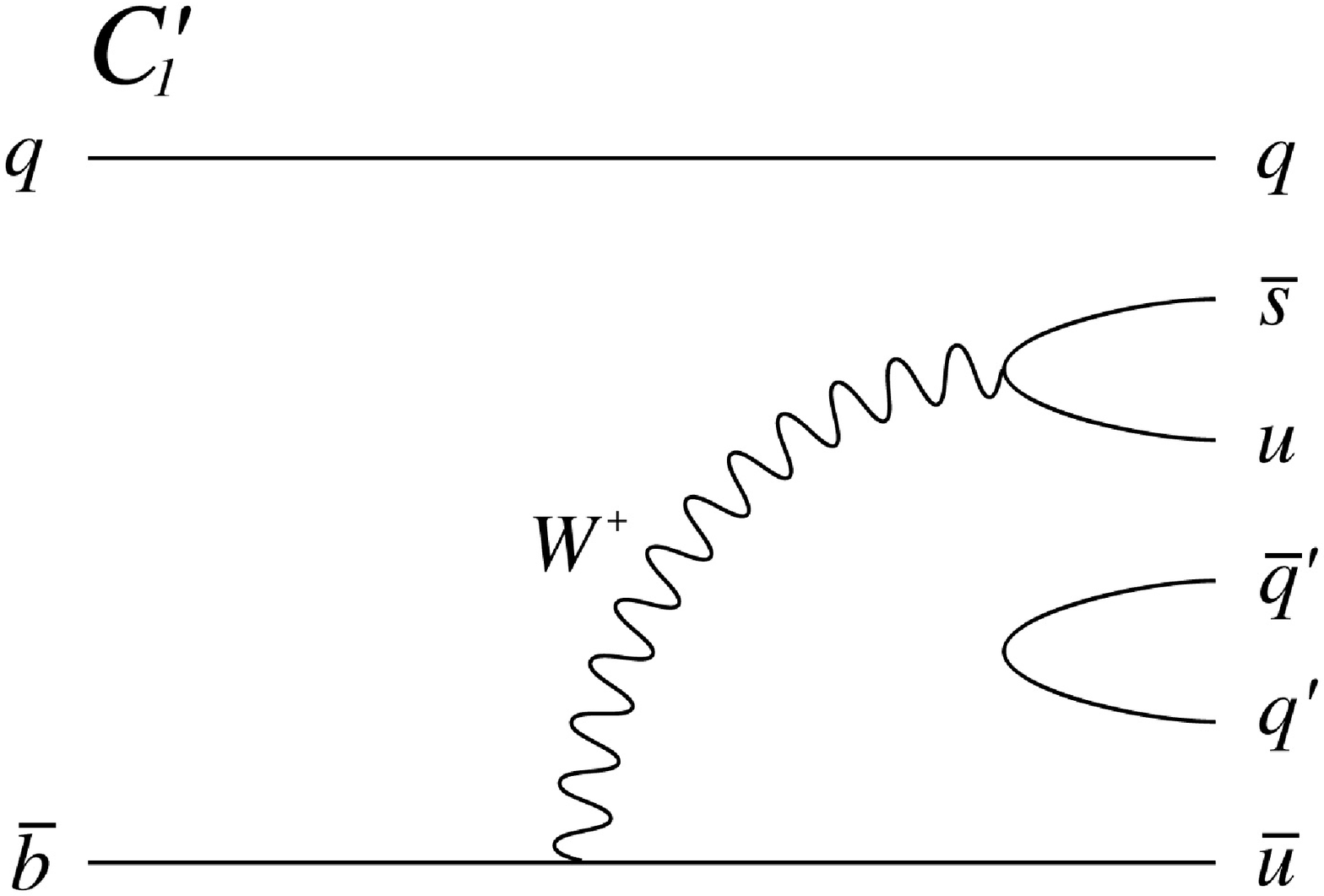}
		\includegraphics[height=3.98cm]{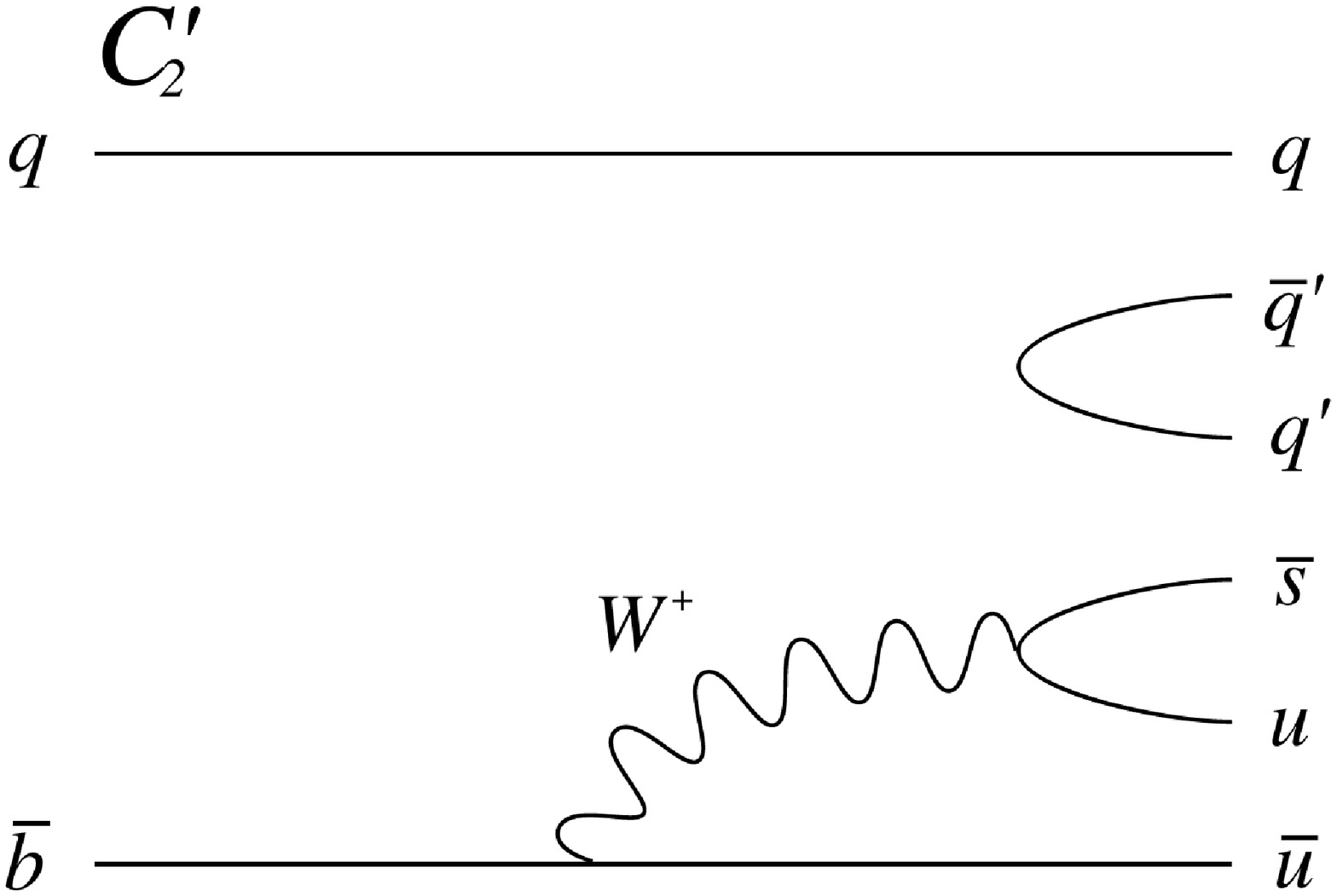}
	\centering
		\includegraphics[height=3.98cm]{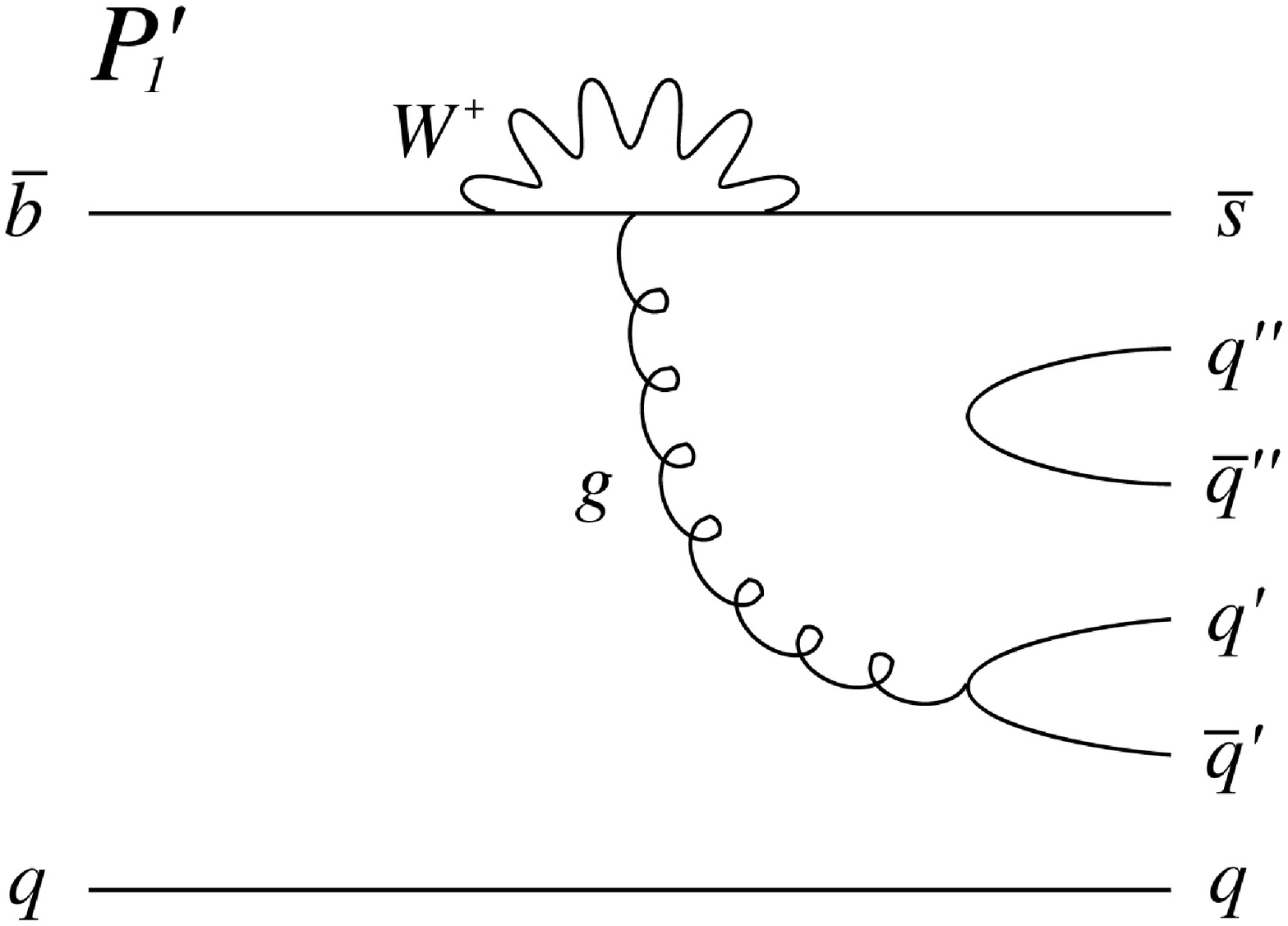}
		\includegraphics[height=3.98cm]{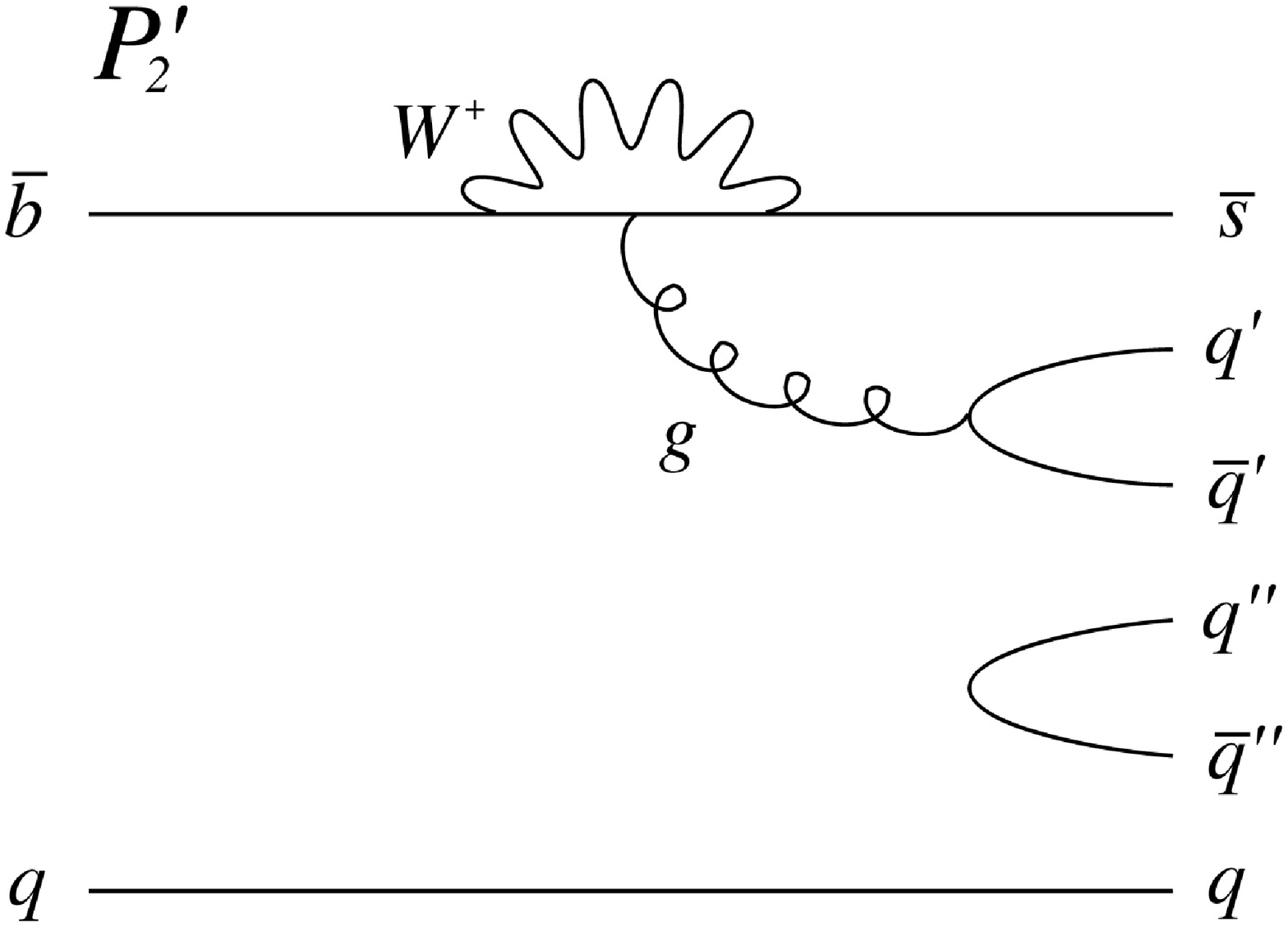}
	\centering
		\includegraphics[height=3.98cm]{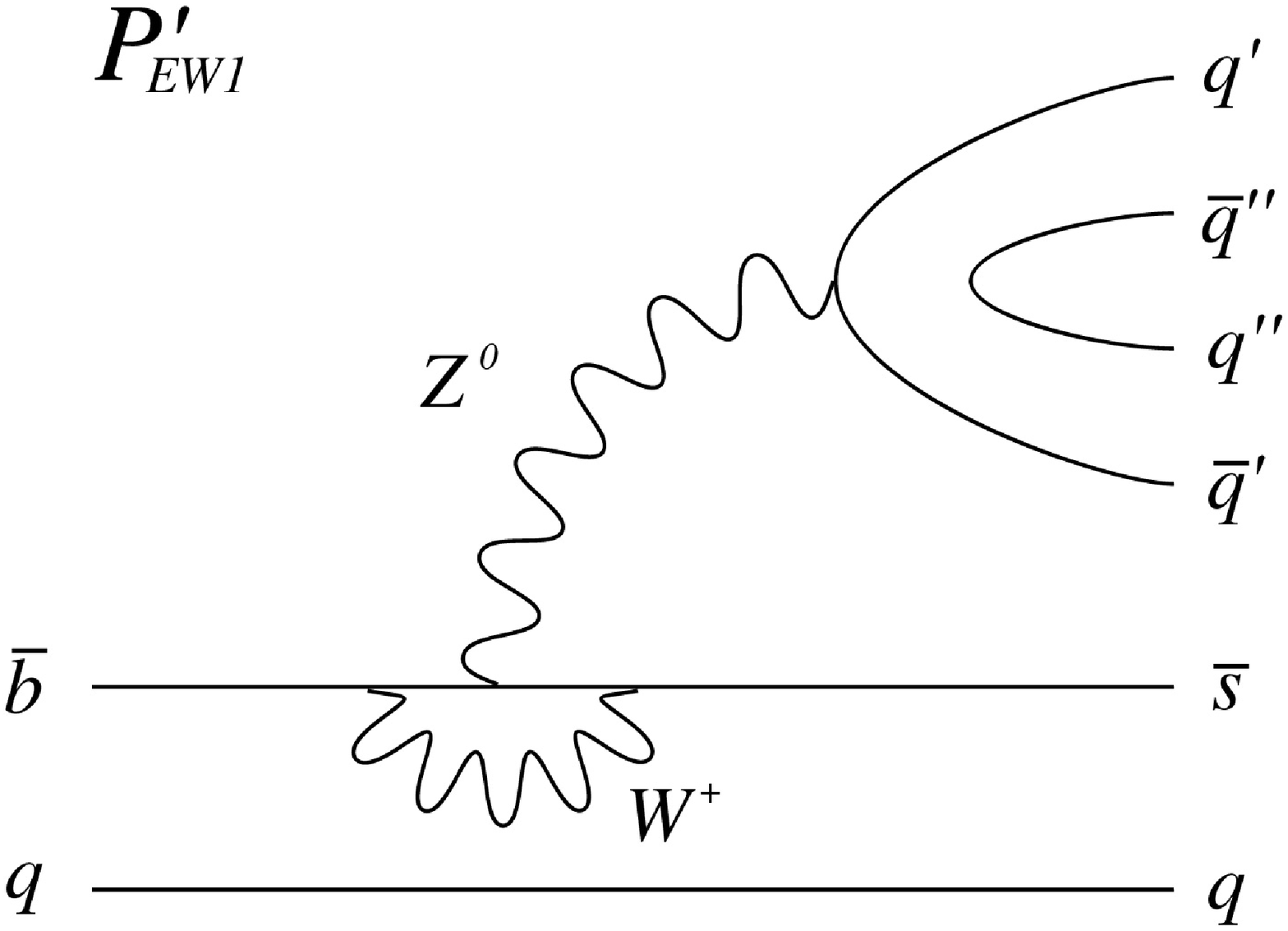}
		\includegraphics[height=3.98cm]{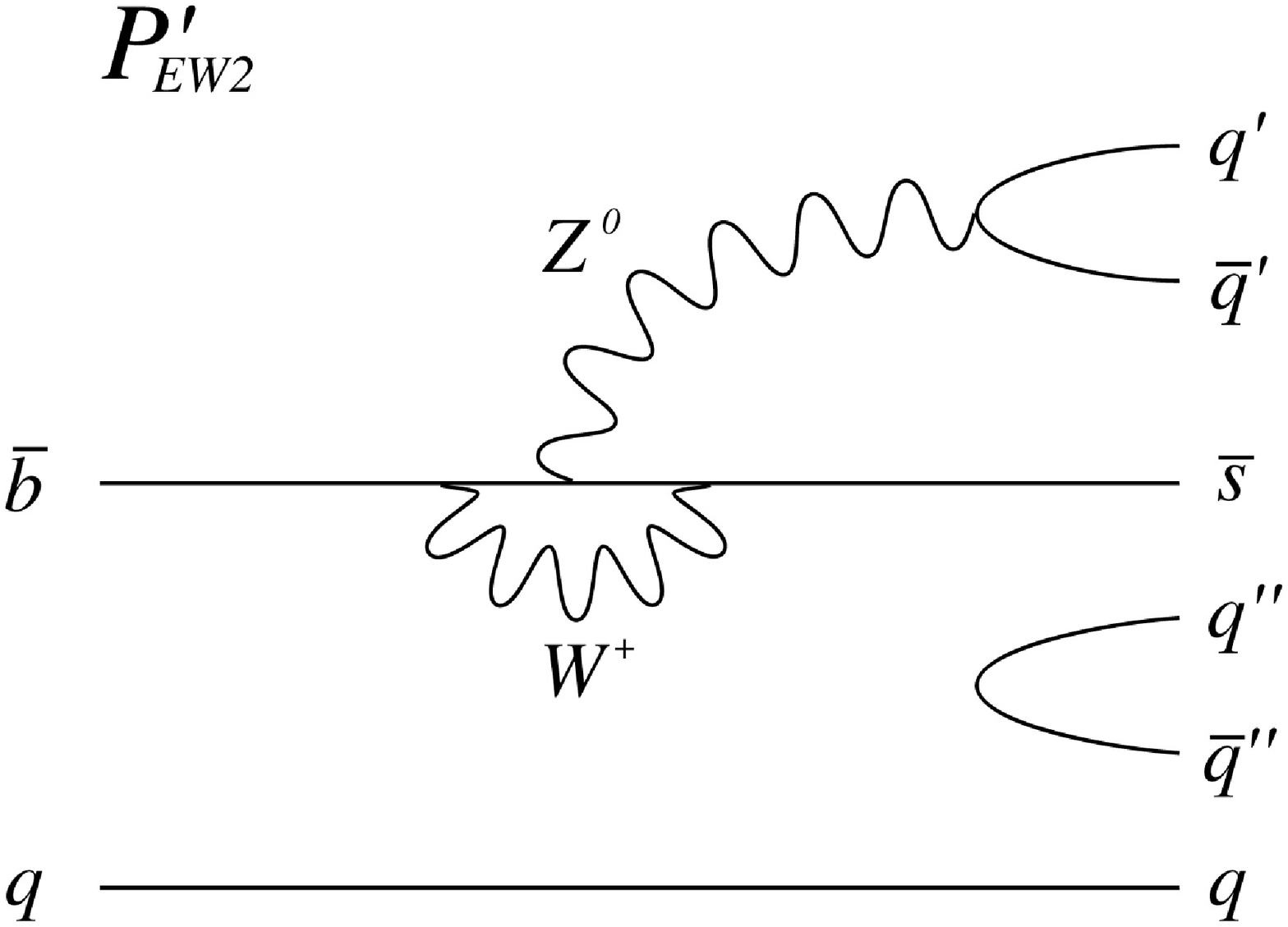}
	\centering
		\includegraphics[height=3.98cm]{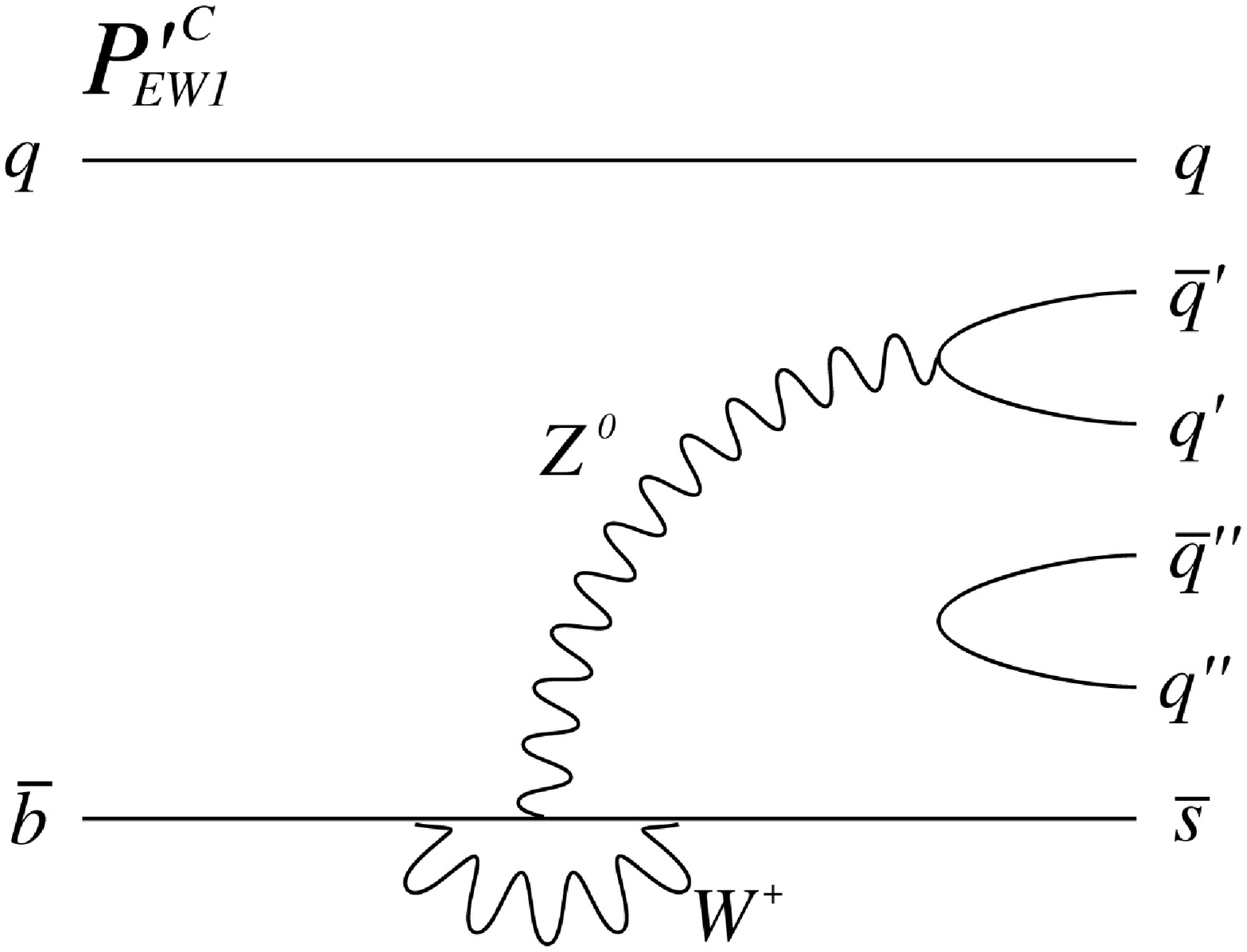}
		\includegraphics[height=3.98cm]{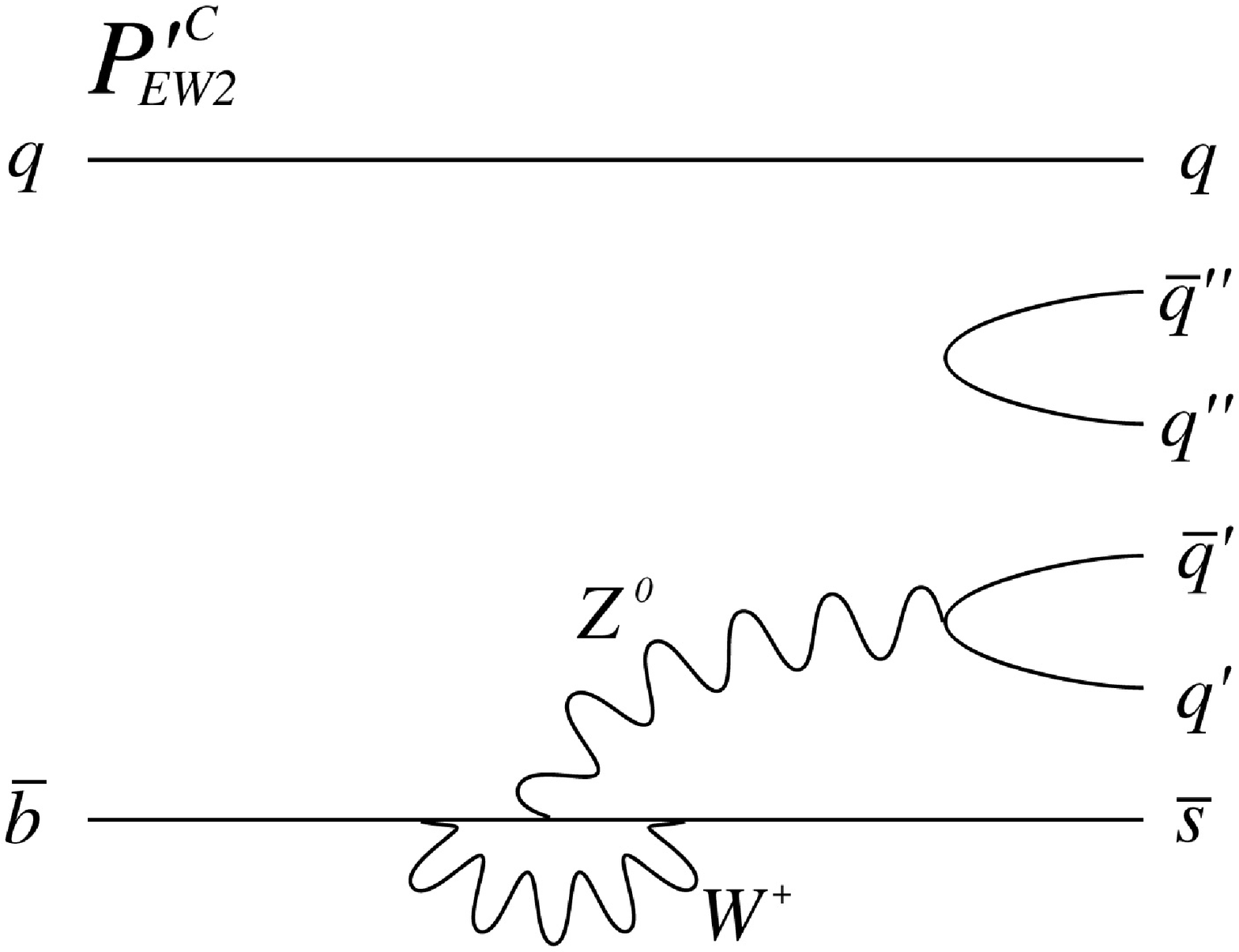}
\caption{Diagrams contributing to $\btokpipi$.}
\label{BKpipifig}
\end{figure}

It was shown in Ref.~\cite{BPPP} that the amplitudes for three-body
$B$ decays can be expressed in terms of diagrams. The diagrams are
shown in Fig.~\ref{BKpipifig} (all annihilation- and exchange-type
diagrams have been neglected). Note:
\begin{itemize}

\item In all diagrams, it is necessary to ``pop'' a quark pair from
  the vacuum. This pair is $u{\bar u}$ or $d{\bar d}$.

\item The subscript ``1'' indicates that the popped quark pair is
  between two (non-spectator) final-state quarks; the subscript ``2''
  indicates that the popped quark pair is between two final-state
  quarks including the spectator.

\end{itemize}
One difference compared to two-body $B$-decays is that here, because
the final state contains three particles, the diagrams are momentum
dependent. However, this does not pose a problem. The diagrams
(magnitudes and relative strong phases) are determined via a fit to
the data. But since the experimental observables are themselves
momentum dependent, the fit will yield the momentum dependence of each
diagram.

In terms of diagrams, the $\btokpipi$ amplitudes are given by
\bea
\sqrt{2} A(B^+ \to K^0\pi^+\pi^0)_{sym} &=& -T'_1 e^{i\gamma}-C'_2 e^{i\gamma} + P'_{EW2} + P^{\prime C}_{EW1} ~, \nn\\
A(\bd \to K^0\pi^+\pi^-)_{sym} &=& -T'_1 e^{i\gamma}-C'_1 e^{i\gamma}-{\tilde P}'_{uc} e^{i\gamma}+ {\tilde P}'_{tc} \nn\\
&& \hskip1.5truecm +~\frac13 P'_{EW1} + \frac23 P^{\prime C}_{EW1} - \frac13 P^{\prime C}_{EW2} ~, \nn\\
\sqrt{2} A(\bd \to K^0\pi^0\pi^0)_{sym} &=& C'_1 e^{i\gamma}- C'_2 e^{i\gamma}+{\tilde P}'_{uc} e^{i\gamma}- {\tilde P}'_{tc}  \nn\\
&& \hskip1.5truecm -~\frac13 P'_{EW1} + P'_{EW2} +\frac13 P^{\prime C}_{EW1} + \frac13 P^{\prime C}_{EW2} ~, \nn\\
A(B^+ \to K^+\pi^+\pi^-)_{sym} &=& -T'_2 e^{i\gamma}-C'_1 e^{i\gamma}-{\tilde P}'_{uc} e^{i\gamma}+ {\tilde P}'_{tc} \nn\\
&& \hskip1.5truecm +~\frac13 P'_{EW1} - \frac13 P^{\prime C}_{EW1} + \frac23 P^{\prime C}_{EW2} ~, \nn\\
\sqrt{2} A(B^+ \to K^+\pi^0\pi^0)_{sym} &=& T'_1 e^{i\gamma}+T'_2 e^{i\gamma}+C'_1 e^{i\gamma}+C'_2 e^{i\gamma}+{\tilde P}'_{uc} e^{i\gamma}- {\tilde P}'_{tc} \nn\\
&& \hskip1.5truecm -~\frac13 P'_{EW1} - P'_{EW2} - \frac23 P^{\prime C}_{EW1} - \frac23 P^{\prime C}_{EW2} ~, \nn\\
\sqrt{2} A(\bd \to K^+\pi^0\pi^-)_{sym} &=& T'_1 e^{i\gamma}+C'_2 e^{i\gamma}- P'_{EW2} - P^{\prime C}_{EW1} ~, 
\label{Kpipisymamps}
\eea
where ${\tilde P}' \equiv P'_1 +P'_2$, and all amplitudes have been
multiplied by $\sqrt{2}$.  Above we have explicitly written the
weak-phase dependence (this includes $\gamma$ and the minus sign from
$V_{tb}^* V_{ts}$ [${\tilde P}'_{tc}$ and EWPs]), while the diagrams
contain strong phases.

Although there are a large number of diagrams in these amplitudes,
they can be combined into a smaller number of effective diagrams:
\bea
\sqrt{2} A(B^+ \to K^0\pi^+\pi^0)_{sym} &=& - T'_a e^{i\gamma} - T'_b e^{i\gamma} + P'_{EW,a} + P'_{EW,b} ~, \nn\\
A(\bd \to K^0\pi^+\pi^-)_{sym} &=& - T'_a e^{i\gamma} - P'_a e^{i\gamma} + P'_b ~, \nn\\
\sqrt{2} A(\bd \to K^0\pi^0\pi^0)_{sym} &=& - T'_b e^{i\gamma} + P'_a e^{i\gamma} - P'_b + P'_{EW,a} + P'_{EW,b} ~, \nn\\
A(B^+ \to K^+\pi^+\pi^-)_{sym} &=& - P'_a e^{i\gamma} + P'_b - P'_{EW,a} ~, \nn\\
\sqrt{2} A(B^+ \to K^+\pi^0\pi^0)_{sym} &=& T'_a e^{i\gamma} + T'_b e^{i\gamma} + P'_a e^{i\gamma} - P'_b - P'_{EW,b} ~, \nn\\
\sqrt{2} A(\bd \to K^+\pi^0\pi^-)_{sym} &=& T'_a e^{i\gamma} + T'_b e^{i\gamma} - P'_{EW,a} - P'_{EW,b} ~,
\label{Kpipieffamps}
\eea
where
\bea
T'_a &\equiv& T'_1 - T'_2 ~,\nn\\
T'_b &\equiv& C'_2 + T'_2 ~,\nn\\
P'_a &\equiv& {\tilde P}'_{uc} + T'_2 + C'_1 ~,\nn\\
P'_b &\equiv& {\tilde P}'_{tc} + \frac13 P'_{EW1} + \frac23 P^{\prime C}_{EW1} - \frac13 P^{\prime C}_{EW2} ~,  \nn\\
P'_{EW,a} &\equiv& P ^{\prime C}_{EW1} - P^{\prime C}_{EW2} ~, \nn\\
P'_{EW,b} &\equiv& P'_{EW2} + P^{\prime C}_{EW2} ~.
\label{eq:effdiag}
\eea
The amplitudes can therefore be written in terms of 6 effective
diagrams. This corresponds to 12 theoretical parameters: 6 magnitudes
of diagrams, 5 relative (strong) phases, and $\gamma$. However, as
noted above, there are only 11 experimental observables. Therefore, in
order to extract $\gamma$, one requires additional input.

One obvious idea is the following. In two-body $\btos$ $B$ decays, the
diagrams are expected to obey the approximate hierarchy \cite{GHLR}
\bea
1 &:& P'_{tc} ~, \nn\\
{\bar\lambda} &:& T', P'_{EW} ~, \nn\\
{\bar\lambda}^2 &:& C', P'_{uc}, P^{\prime C}_{EW}  ~,
\eea
where ${\bar\lambda} \simeq 0.2$. If the three-body decay diagrams
obey a similar hierarchy, one can neglect $C'_1$, $C'_2$, ${\tilde
  P}'_{uc}$, $P^{\prime C}_{EW1}$, $P^{\prime C}_{EW2}$, with only a
$\sim 5\%$ theoretical error. But if these diagrams are neglected,
then two of the effective diagrams vanish: $P'_{EW,a} \to 0$ and $T'_b
- P'_a \to 0$ [Eq.~(\ref{eq:effdiag})]. In this case, the amplitudes
can be written in terms of 4 effective diagrams, corresponding to 8
theoretical parameters: 4 magnitudes of diagrams, 3 relative (strong)
phases, and $\gamma$.  Given that there are 11 experimental
observables, the weak phase $\gamma$ can be extracted.

The problem here is that it is difficult to test the assumption that
$C'_1$, $C'_2$, ${\tilde P}'_{uc}$, $P^{\prime C}_{EW1}$ and
$P^{\prime C}_{EW2}$ are negligible, so that the theoretical error is
really unknown. Given this, it is perhaps better to look for another
method, in which the theoretical error is better under control.

As mentioned in the introduction, in Ref.~\cite{DSS}, Deshpande, Sinha
and Sinha (DSS) proposed a new method for measuring $\gamma$ in $B\to
K\pi\pi$ decays. Although the details are different, at its heart the
method is similar to that outlined above. While DSS do not write the
amplitudes in terms of diagrams, they do note that each decay
amplitude receives two contributions, one proportional to
$e^{i\gamma}$, the other with no weak phase. The key point is that
there is no gluonic-penguin contribution to $B^+ \to K^0\pi^+\pi^0$ --
its amplitude has only tree and EWP pieces. DSS' assumption, which
provides the additional input and allows $\gamma$ to be extracted, is
that the EWP and tree contributions in $B^+ \to K^0\pi^+\pi^0$ are
related to one another as in Refs.~\cite{NR,GPY}. Unfortunately, it was
then shown that this relation does not hold \cite{Grocomment}, so that
$\gamma$ cannot be obtained using DSS' method.

Now, in terms of diagrams, the DSS assumption is that $T'_1 + C'_2$ is
related to $P'_{EW2} + P^{\prime C}_{EW1}$ [Eq.~(\ref{Kpipisymamps})].
Although this is not true, it does not preclude other EWP-tree
relations. Indeed, as we will see in the next section, such relations
do exist, and their imposition does allow $\gamma$ to be extracted
from $\btokpipi$ decays.

Finally, we return to the issue of the underlying symmetry. The above
discussion is for the case where only isospin symmetry is considered.
However, below we will see that it may be necessary to assume full
flavor SU(3) symmetry. In this case, the final state involves three
identical particles, so that the six permutations of these particles
(the group $S_3$) must be taken into account. Correspondingly, there
are six possible wavefunctions, in which the three particles are in a
totally symmetric state, a totally antisymmetric state, or one of four
mixed states. These six states can be chosen such that the $\pi\pi$
wavefunction is either symmetric or antisymmetric. A symmetric
$\pi\pi$ state is then a linear combination of the totally symmetric
$S_3$ state and one mixed state. Consequently, the parametrization of
Eq.~(\ref{Kpipieffamps}) holds even under full SU(3) symmetry, as long
the state is symmetric under $\pi\pi$ exchange.

\section{EWP-tree Relations}

EWP-tree relations are well known in the context of $B \to PP$ decays
($P$ is a pseudoscalar meson), particularly $\btokpi$.  They have
been very useful for reducing the number of free theoretical
parameters. The starting point is the electroweak effective
hamiltonian for quark-level ${\bar b}$ decays \cite{BBL}:
\beq 
H_{eff} = {G_F \over \sqrt{2}} \sum_{q=d,s}
\left(\sum_{p=u,c}\lambda_p^{(q)} (c_1(\mu) O^p_1 (\mu) + c_2(\mu)
O^p_2 (\mu)) - \lambda_t^{(q)} \sum_{i=3}^{10} c_i(\mu) O_i(\mu)
\right) ~,
\label{Heff}
\eeq
where $\lambda_p^{(q)}=V^*_{pb} V_{pq}$. $\mu$ is the renormalization
point, typically taken to be $O(m_b)$. All physical quantities must be
independent of $\mu$.  The Wilson coefficients $c_i$ include
gluons (QCD corrections) whose energy is above $\mu$ (short distance),
while the operators $O_i$ include QCD corrections of energy less than
$\mu$ (long distance). Note: factors of $G_F/\sqrt{2}$ are omitted for
the remainder of this paper.

The operators take the following form:
\beq
  O_1^p = (\bar b_\alpha p_\alpha)_{V-A}\,
  (\bar p_\beta q_\beta)_{V-A} ~~,~~~~
  O_2^p = (\bar b_\alpha p_\beta)_{V-A}\,
  (\bar p_\beta q_\alpha)_{V-A} ~,
\eeq
summed over color indices $\alpha$ and $\beta$. These are the usual
(tree-level) current-current operators induced by $W$-boson exchange.
\bea
  O_3 = (\bar b_\alpha q_\alpha)_{V-A}\,\sum_{q'}\,
  (\bar q'_\beta q'_\beta)_{V-A} & , &
  O_4 = (\bar b_\alpha q_\beta)_{V-A}\,\sum_{q'}\,
  (\bar q'_\beta q'_\alpha)_{V-A} \,, \nonumber\\
  O_5 = (\bar b_\alpha q_\alpha)_{V-A}\,\sum_{q'}\,
  (\bar q'_\beta q'_\beta)_{V+A} & , &
  O_6 = (\bar b_\alpha q_\beta)_{V-A}\,\sum_{q'}\,
  (\bar q'_\beta q'_\alpha)_{V+A} \,,
\eea
summed over the light flavors $q'=u,d,s,c$. These are referred to as
QCD (gluonic) penguin operators.
\bea
  O_7 = \frac32\,(\bar b_\alpha q_\alpha)_{V-A}\,\sum_{q'}\,e_{q'}\,
  (\bar q'_\beta q'_\beta)_{V+A} & , &
  O_8 = \frac32\,(\bar b_\alpha q_\beta)_{V-A}\,\sum_{q'}\,e_{q'}\,
  (\bar q'_\beta q'_\alpha)_{V+A} \,, \nonumber\\
  O_9 = \frac32\,(\bar b_\alpha q_\alpha)_{V-A}\,\sum_{q'}\,e_{q'}\,
  (\bar q'_\beta q'_\beta)_{V-A} & , &
  O_{10} = \frac32\,(\bar b_\alpha q_\beta)_{V-A}\,\sum_{q'}\,e_{q'}\,
  (\bar q'_\beta q'_\alpha)_{V-A} \,,
\label{EWPops}
\eea
with $e_{q'}$ denoting the electric charges of the quarks. These are
the electroweak-penguin operators. The quark current $(\bar
q_1 q_2)_{V\pm A}$ denotes $\bar q_1\gamma^\mu (1\pm\gamma_5)q_2$.
The key observation is that the Wilson coefficients $c_{7,8}$ are
small compared to $c_{9,10}$.  Neglecting them, the tree and EWP
operators then have exactly the same structure, up to a Fierz
transformation of the fermions, and can be related.

Various approaches have been used to exploit this fact for $\btokpi$
decays.  Neubert and Rosner (NR) showed that a basic SU(3) EWP-tree
relation can be obtained by manipulating the effective hamiltonian
itself at the level of quark operators \cite{NR}.  Later, Gronau,
Pirjol and Yan (GPY) used a more general technique based on group
theory to find additional SU(3) EWP-tree relations \cite{GPY}.
Recently, it was shown that these relations can be obtained by
studying Wick contractions of the effective hamiltonian
\cite{contractions}.

In this section, we will apply the contractions approach to $B \to
K\pi\pi$ decays. As we will see, the correct SU(3) EWP-tree relations
in $B\to K \pi \pi$ are between specific diagrams.  For example,
$P'_{EW1}$ is related to $T'_1$ and $C'_1$, and not to $T'_2$ and
$C'_2$. Since different diagrams such as $T'_1$ and $T'_2$ cannot be
distinguished at the level of operators or group theory, the NR and
GPY approaches may not be applicable \cite{KpipiSU3}.  In the
following subsection, we give a brief review of the contractions
formalism.

\subsection{Contractions}

The formalism of contractions gives a bridge between the effective
hamiltonian and the language of diagrams. Contractions include all the
short-distance information of Wilson coefficients, and also exploit
the fact that trees and EWPs arise from long-distance operators with
almost identical structures. In Ref.~\cite{contractions}, contractions
are discussed at length for $B \to PP$ decays (see also
Ref.~\cite{BS}). Here only isospin symmetry is assumed initially. It
is shown that all diagrams can be expressed in terms of contractions,
and the EWP-tree relations of Refs.~\cite{NR,GPY} are reproduced.
However, these relations hold only if SU(3) symmetry is imposed. For
this reason, in our review below, we assume SU(3) from the beginning.
Also, for definitiveness, and to make the comparison with $\btokpipi$
clearer, we focus on the decay $\btokpi$.

The idea is as follows: (i) one symmetrizes or antisymmetrizes the
final state, (ii) one takes the operators of effective hamiltonian,
(iii) one adds initial and final states, and (iv) one computes the sum
of all possible Wick contractions, applying the basic rules of quantum
field theory. This gives the decomposition of the decay amplitude in
terms of contractions. This can be compared with the decomposition in
terms of diagrams, and therefore gives us the structure of each
diagram in terms of contractions. It is this comparison which allows
us to match diagrams and contractions, and thus yields the EWP-tree
relations.

Since the spinless $B$ meson decays into a pair of pseudoscalar mesons
$K$ and $\pi$, these are necessarily in an $S$-wave.  Under SU(3), $K$
and $\pi$ mesons are identical particles, and so one must symmetrize
the final state $\ket{f}$:
\beq
\ket{f} = \frac{1}{\sqrt{2}} \left(  \ket{K(p_1) \pi (p_2)} + \ket{\pi(p_1) K(p_2)}\right)~.
\label{eq:symPP}
\eeq
When calculating the amplitude for a particular $B \to K \pi$ decay,
one must ``sandwich'' all operators of the effective Hamiltonian
between initial and final states. All such terms have the form
\beq
\langle \bar q_1 q_2 \bar q_3 q_4 | \bar b q_5 \, \bar q_6 q_7 |
\bar b q_8 \rangle~.
\label{contracform}
\eeq
(Dirac and color structures are omitted for notational convenience.)
$\bar b q_8$ is the $B$ meson. The final-state mesons contain the
quarks $\bar q_1$, $q_2$, $\bar q_3$ and $q_4$. The two choices are $K
= \bar q_1 q_2$ and $\pi =\bar q_3 q_4$, or $\pi = \bar q_1 q_2$ and
$K =\bar q_3 q_4$, and these correspond to the two states in
Eq.~(\ref{eq:symPP}).

For a given $B$ decay, there are $4! = 24$ possible contractions.
However, not all are independent. For example, consider the two
contractions\footnote{Here `$EM$' stands for ``emission.'' See
  Ref.~\cite{contractions} for details.}
\beq
EM'(1)=
{
\contraction[1.4ex]{\langle \bar q_1 q_2 \bar q_3 }{q_4 }{| \bar b q_5 }{\bar q_6 }
\contraction[2ex]{\langle \bar q_1 q_2 }{\bar q_3 }{q_4 | \bar b q_5 \bar q_6 }{q_7 }
\contraction[4ex]{\langle }{\bar q_1 }{q_2 \bar q_3 q_4 | \bar b }{q_5 }
\contraction[3.4ex]{\langle \bar q_1 }{q_2 }{\bar q_3 q_4 | \bar b q_5 \bar q_6 q_7 | \bar b }{q_8 }
\langle \bar q_1 q_2 \bar q_3 q_4 | \bar b q_5 \bar q_6 q_7 | \bar b q_8 \rangle
}~,~~
EM'(2)=
{
\contraction[1ex]{\langle \bar q_1 q_2 }{\bar q_3 }{q_4 | \bar b }{q_5 }
\contraction[2.4ex]{\langle \bar q_1 }{q_2 }{\bar q_3 q_4 | \bar b q_5 }{\bar q_6 }
\contraction[3ex]{\langle }{\bar q_1 }{q_2 \bar q_3 q_4 | \bar b q_5 \bar q_6 }{q_7 }
\contraction[4.4ex]{\langle \bar q_1 q_2 \bar q_3 }{q_4 }{| \bar b q_5 \bar q_6 q_7 | \bar b }{q_8 }
\langle \bar q_1 q_2 \bar q_3 q_4 | \bar b q_5 \bar q_6 q_7 | \bar b q_8 \rangle
}~.
\label{tageqn}
\eeq
(The prime indicates a $\btos$ transition.)  Here the labels $(1)$ and
$(2)$ correspond respectively to the momentum assignments $K(p_1)
\pi(p_2)$ and $\pi(p_1) K(p_2)$. It is clear that the above
contractions are not independent since one can be obtained from the
other with an exchange of mesons, so that $EM'(1)=EM'(2)$.

Now, if one performs the contractions with the operators $O_1^u$ and
$O_2^u$ of Eq.~(\ref{Heff}), one finds that the $T'$ diagram is
related to the $EM'$-type contractions \cite{contractions}:
\bea
T' &=& \frac{1}{\sqrt{2}}  \, |\lambda_u^{(s)}| ( c_1 EM'_1 (1) + c_1 EM'_1 (2) + c_2 EM'_2 (1) + c_2 EM'_2 (2) )\nn\\ 
&=& \frac{1}{\sqrt{2}}  \, |\lambda_u^{(s)}|  \, c_1 \left( EM'_1(1) + EM'_1(2) + {c_2 \over c_1} EM'_2(1)+ {c_2 \over c_1} EM'_2(2) \right) ~,
\eea
where $EM'_i$ is an $EM'$-type contraction of the operator $O_i$.
Similarly, the $\pewp$ diagram is related to the $EM'_i$ contraction
of the operators $O_9$ and $O_{10}$:
\bea
\pewp &\!\!=\!\!&  -\frac{1}{\sqrt{2}} \, \frac{3}{2}  \, |\lambda_t^{(s)}| ( c_9 EM'_9(1) + c_9 EM'_9(2) + c_{10} EM'_{10}(1)+ c_{10} EM'_{10}(2) )\nn\\
 &\!\!=\!\!& -\frac{1}{\sqrt{2}} \, \frac{3}{2}  \, |\lambda_t^{(s)}|  \,  c_9 \left( EM'_9(1) + EM'_9(2) + {c_2 \over c_1} EM'_{10}(1) + {c_2 \over c_1} EM'_{10}(2)  \right) ~,
\eea
Here, we have used the fact that the Wilson coefficients obey $c_1/c_2
= c_9/c_{10}$ to about 5\%. (In the rest of the paper, we assume this
equality.)

Now, the $T'$ diagram contains $EM'$-type contractions of $O_{1,2}^u$,
while the $\pewp$ diagram contains $EM'$-type contractions of
$O_{9,10}$.  However, since $s$-quark contractions are equal to $u$-
or $d$-quark contractions in the SU(3) limit, $O_9^q \sim (\bar
b_\alpha s_\alpha)_{V-A} (\bar q_\beta q_\beta)_{V-A} = (\bar b_\alpha
u_\alpha)_{V-A}\, (\bar u_\beta s_\beta)_{V-A} \sim O_1^u$. That is,
$O_9^q$ and $O_1^u$ have the same form under SU(3). Things are similar
for $O_{10}^q$ and $O_2^u$.  We therefore see that $\pewp$ is
proportional to $T'$:
\beq
\pewp = -\frac{3}{2} \, \frac{|\lambda_t^{(s)}|}{|\lambda_u^{(s)}|} \, \frac{c_9+c_{10}}{c_1+c_2} \, T' ~.
\label{EWPtreerel1}
\eeq

The argument is much the same for $C'$ and $\pewcp$. Two other
contractions are
\beq
EM'_{C}(1)=
{
\contraction[1ex]{\langle }{\bar q_1 }{q_2 \bar q_3 q_4 | \bar b }{q_5 }
\contraction[2.4ex]{\langle \bar q_1 }{q_2 }{\bar q_3 q_4 | \bar b q_5 }{\bar q_6 }
\contraction[3ex]{\langle \bar q_1 q_2 }{\bar q_3 }{q_4 | \bar b q_5 \bar q_6 }{q_7 }
\contraction[4.4ex]{\langle \bar q_1 q_2 \bar q_3 }{q_4 }{| \bar b q_5 \bar q_6 q_7 | \bar b }{q_8 }
\langle \bar q_1 q_2 \bar q_3 q_4 | \bar b q_5 \bar q_6 q_7 | \bar b q_8 \rangle
}~,~~
EM'_{C}(2)=
{
\contraction[1ex]{\langle \bar q_1 q_2 }{\bar q_3 }{q_4 | \bar b }{q_5 }
\contraction[2.4ex]{\langle \bar q_1 q_2 \bar q_3 }{q_4 }{| \bar b q_5 }{\bar q_6 }
\contraction[3ex]{\langle }{\bar q_1 }{q_2 \bar q_3 q_4 | \bar b q_5 \bar q_6 }{q_7 }
\contraction[4.4ex]{\langle \bar q_1 }{q_2 }{\bar q_3 q_4 | \bar b q_5 \bar q_6 q_7 | \bar b }{q_8 }
\langle \bar q_1 q_2 \bar q_3 q_4 | \bar b q_5 \bar q_6 q_7 | \bar b q_8 \rangle
}~.
\eeq
The diagrams $C'$ and $\pewcp$ are related to the $EM'_{C}$-type contractions:
\bea
C' &=& \frac{1}{\sqrt{2}}  \, |\lambda_u^{(s)}| ( c_1 EM'_{{C}1}(1) + c_1 EM'_{{C}1}(2) + c_2 EM'_{{C}2}(1) + c_2 EM'_{{C}2}(2) )\nn\\
&=&\frac{1}{\sqrt{2}}  \, |\lambda_u^{(s)}|  \, c_1 \left( EM'_{{C}1}(1) + EM'_{{C}1}(2) + {c_2 \over c_1} EM'_{{C}2}(1)+ {c_2 \over c_1} EM'_{{C}2}(2) \right) ~, \\
\pewcp &=&  -\frac{1}{\sqrt{2}} \, \frac{3}{2}  \, |\lambda_t^{(s)}| ( c_9 EM'_{{C}9}(1) + c_9 EM'_{{C}9}(2) + c_{10} EM'_{{C}10}(1) + c_{10} EM'_{{C}10}(2) )\nn\\ 
&=& -\frac{1}{\sqrt{2}} \, \frac{3}{2}  \, |\lambda_t^{(s)}|  \,  c_9 \left( EM'_{{C}9}(1) + EM'_{{C}9}(2) + {c_2 \over c_1} EM'_{{C}10}(1) + {c_2 \over c_1} EM'_{{C}10}(2) \right) ~. \nn
\eea
In the SU(3) limit, $EM'_{{C}9}(n) = EM'_{{C}1}(n)$ and
$EM'_{{C}10}(n) = EM'_{{C}2}(n)$ ($n=1,2$), so that $\pewcp$ is
proportional to $C'$:
\beq
\pewcp = -\frac{3}{2} \, \frac{|\lambda_t^{(s)}|}{|\lambda_u^{(s)}|} \, \frac{c_9+c_{10}}{c_1+c_2} \, C' ~.
\label{EWPtreerel2}
\eeq

Above, we have described the formalism of contractions in the context
of two-body decays.  Our aim now is to apply this to the problem of $B
\to K \pi \pi$ decays, and derive EWP-tree relations. As we saw above,
different contractions can be made equal through the imposition of
SU(3). However, this can lead to some subtleties in the case of
three-body decays.

Under SU(3), $\pi$ and $K$ mesons are treated as identical particles,
and the total wavefunction of the final state must be symmetric under
the exchange of these particles. For $\btokpi$ decays, since the final
state has to be in an $S$-wave, it is automatically symmetric under
the exchange of the final-state mesons.  However, for $B \to K \pi
\pi$, higher states of angular momentum are possible, and the final
state is then not necessarily symmetric under permutations of the
mesons. As was mentioned earlier, the group of permutations is $S_3$,
and there are six possible states: the three particles can be in a
totally symmetric state, a totally antisymmetric state, or one of four
mixed states. To be completely explicit, we define
\bea
\ket{1} &\equiv& \ket{K(p_1) \pi_1(p_2) \pi_2(p_3)}~,\nn\\ 
\ket{2} &\equiv& \ket{K(p_1) \pi_2(p_2) \pi_1(p_3)}~,\nn\\ 
\ket{3} &\equiv& \ket{\pi_2(p_1) K(p_2) \pi_1(p_3)}~,\nn\\ 
\ket{4} &\equiv& \ket{\pi_2(p_1) \pi_1(p_2) K(p_3)}~,\nn\\ 
\ket{5} &\equiv& \ket{\pi_1(p_1) \pi_2(p_2) K(p_3)}~,\nn\\ 
\ket{6} &\equiv& \ket{\pi_1(p_1) K(p_2) \pi_2(p_3)}~,
\eea
where the $p_i$ are the momenta of the final-state mesons. The six
states of $S_3$ can then be defined as
\bea
\ket{S} &\equiv& \frac{1}{\sqrt{6}} \left( \ket{1} + \ket{2} + \ket{3} + \ket{4} + \ket{5} + \ket{6} \right)~,\nn\\
\ket{M_1} &\equiv& \frac{1}{\sqrt{12}} \left( 2\ket{1} + 2\ket{2} - \ket{3} - \ket{4} - \ket{5} - \ket{6} \right)~,\nn\\
\ket{M_2} &\equiv& \frac{1}{\sqrt{4}} \left( \ket{3} - \ket{4} - \ket{5} + \ket{6} \right)~,\nn\\
\ket{M_3} &\equiv& \frac{1}{\sqrt{4}} \left( -\ket{3} - \ket{4} + \ket{5} + \ket{6} \right)~,\nn\\
\ket{M_4} &\equiv& \frac{1}{\sqrt{12}} \left( 2\ket{1} - 2\ket{2} - \ket{3} + \ket{4} - \ket{5} + \ket{6} \right)~,\nn\\
\ket{A} &\equiv& \frac{1}{\sqrt{6}} \left( \ket{1} - \ket{2} + \ket{3} - \ket{4} + \ket{5} - \ket{6} \right)~.
\label{SU3states}
\eea
Note that $\ket{S}$, $\ket{M_1}$ and $\ket{M_2}$ are all symmetric
under the exchange of the two pions, while $\ket{M_3}$,
$\ket{M_4}$ and $\ket{A}$ are all antisymmetric.

Below, we present two cases which illustrate the features of all six
$S_3$ states. First, we examine the totally symmetric SU(3) state
$\ket{S}$. This can be determined experimentally as follows. Consider
again the decay $\bd \to K^0\pi^+\pi^-$. In Sec.~\ref{amps} it was
noted that the Dalitz-plot events can be described by $s_+$ and $s_-$
[Eq.~(\ref{s+s-defs})], and that the decay amplitude, ${\cal
  M}(s_+,s_-)$, can be extracted. We introduce the third Mandelstam
variable, $s_0 = \left( p_{\pi^+} + p_{\pi^-} \right)^2$. It is
related to $s_+$ and $s_-$ as follows:
\beq
s_0 = m_B^2 + 2m_\pi^2 + m_{K^0}^2 - s_+ - s_- ~.
\eeq
The totally symmetric SU(3) decay amplitude is then given by
\beq
\frac{1}{\sqrt{6}} \left[ {\cal M}(s_+,s_-) + {\cal M}(s_-,s_+) +
  {\cal M}(s_+,s_0) + {\cal M}(s_0,s_+) + {\cal M}(s_0,s_-) + {\cal
    M}(s_-,s_0) \right] ~.
\eeq
Other decays can be treated similarly.

Second, we examine the state which is symmetric only under the
exchange of the two pions (we denote this state as $\ket{S_{\pi\pi}}$,
and refer to it as $\pi\pi$-symmetric).  Previous analyses of
$\btokpipi$ concentrated on the $\pi\pi$-symmetric case with isospin
symmetry \cite{LNQS,DSS,BPPP,GR2005}. It is written as
\bea
\ket{S_{\pi\pi}} &=& \frac{1}{\sqrt{2}} \left(\ket{K (p_1) \pi_1 (p_2) \pi_2 (p_3)} + \ket{K (p_1) \pi_2 (p_2) \pi_1 (p_3)}\right) \nn\\
&=& \sqrt{\frac{1}{3}} \ket{S} + \sqrt{\frac{2}{3}} \ket{M_1} ~.
\label{eq:pipisymstate}
\eea
Thus, the $\pi\pi$-symmetric state is a mixture of the totally
symmetric state and a mixed state of $S_3$. 

\subsection{Totally symmetric case}

We begin with the totally symmetric state $\ket{S}$. The amplitude is
obtained by summing over all possible contractions:
\beq
\mathcal{A}(B \to K \pi_1 \pi_2)_{tot{\hbox{-}}sym} = \sum_{contractions} \bra{S} \mathcal{H}_{eff} \ket{B}~.
\eeq
Here there are $5! = 120$ possible contractions. Even after removing
those which are not independent, there are a large number of
contractions involved.

Below we concentrate on the tree and EWP contractions/diagrams. We use
the same notation as for $\btokpi$.  To be specific, $X_i(n)$
($n=1$-6) is an $X$-type contraction of the operator $O_i$ of
$\mathcal{H}_{eff}$ arising from the momentum assignments of the
states $\ket{n} = \ket{1}$, $\ket{2}$, ..., $\ket{6}$. For example,
$T'_{1,2}(2)$ denotes a contraction of the tree operator $O_2$ related
to the $T'_1$ diagram with momentum assignments $K(p_1)$, $\pi_1(p_3)$
and $\pi_2(p_2)$. The explicit forms of contractions for the trees and
EWPs that interest us are the following:
\bea
C'_1(1)=
{
\contraction[1ex]{\langle \bar q_1 q_2 }{\bar q_3 }{q_4 \bar q_5}{ q_6 }
\contraction[2ex]{\langle \bar q_1 q_2 \bar q_3 q_4 }{\bar q_5 }{q_6 | \bar b }{q_7 }
\contraction[3.5ex]{\langle \bar q_1 q_2 \bar q_3 }{q_4 }{\bar q_5 q_6 | \bar b q_7 }{\bar q_8 }
\contraction[4.5ex]{\langle \bar q_1 }{q_2 }{\bar q_3 q_4 \bar q_5 q_6 | \bar b q_7 \bar q_8 q_9 | \bar b }{q_{10}}
\contraction[5ex]{\langle }{\bar q_1 }{q_2 \bar q_3 q_4 \bar q_5 q_6 | \bar b q_7 \bar q_8 }{q_9 }
\langle \bar q_1 q_2 \bar q_3 q_4 \bar q_5 q_6 | \bar b q_7 \bar q_8 q_9 | \bar b q_{10} \rangle
}
~&,&~~
T'_1(1)=
{
\contraction[1.4ex]{\langle \bar q_1 }{q_2 }{}{\bar q_3 }
\contraction[2.9ex]{\langle }{\bar q_1 }{q_2 \bar q_3 q_4 \bar q_5 q_6 | \bar b q_7 \bar q_8 }{q_9 }
\contraction[2.4ex]{\langle \bar q_1 q_2 \bar q_3 }{q_4 }{\bar q_5 q_6 | \bar b q_7 }{\bar q_8 }
\contraction[4ex]{\langle \bar q_1 q_2 \bar q_3 q_4 \bar q_5 }{q_6 }{| \bar b q_7 \bar q_8 q_9 | \bar b }{q_{10} }
\contraction[1ex]{\langle \bar q_1 q_2 \bar q_3 q_4 }{\bar q_5 }{q_6 | \bar b }{q_7 }
\langle \bar q_1 q_2 \bar q_3 q_4 \bar q_5 q_6 | \bar b q_7 \bar q_8 q_9 | \bar b q_{10} \rangle
}
~,\nn\\
C'_2(1)=
{
\contraction[1ex]{\langle \bar q_1 }{q_2 }{}{\bar q_3 }
\contraction[2.6ex]{\langle }{\bar q_1 }{q_2 \bar q_3 q_4 \bar q_5 q_6 | \bar b q_7 \bar q_8 }{q_9 }
\contraction[3.7ex]{\langle \bar q_1 q_2 \bar q_3 }{q_4 }{\bar q_5 q_6 | \bar b q_7 \bar q_8 q_9 | \bar b }{q_{10} }
\contraction[2ex]{\langle \bar q_1 q_2 \bar q_3 q_4 \bar q_5 }{q_6 }{| \bar b q_7 }{\bar q_8 }
\contraction[1ex]{\langle \bar q_1 q_2 \bar q_3 q_4 }{\bar q_5 }{q_6 | \bar b }{q_7 }
\langle \bar q_1 q_2 \bar q_3 q_4 \bar q_5 q_6 | \bar b q_7 \bar q_8 q_9 | \bar b q_{10} \rangle
}
~&,&~~
T'_2(1)=
{
\contraction[4.2ex]{\langle \bar q_1 }{q_2 }{\bar q_3 q_4 \bar q_5 q_6 | \bar b q_7 }{\bar q_8 }
\contraction[5ex]{\langle }{\bar q_1 }{q_2 \bar q_3 q_4 \bar q_5 q_6 | \bar b q_7 \bar q_8 }{q_9 }
\contraction[3.2ex]{\langle \bar q_1 q_2 \bar q_3 }{q_4 }{\bar q_5 q_6 | \bar b q_7 \bar q_8 q_9 | \bar b }{q_{10} }
\contraction[1ex]{\langle \bar q_1 q_2 }{\bar q_3 }{q_4 \bar q_5 }{q_6 }
\contraction[2ex]{\langle \bar q_1 q_2 \bar q_3 q_4 }{\bar q_5 }{q_6 | \bar b }{q_7 }
\langle \bar q_1 q_2 \bar q_3 q_4 \bar q_5 q_6 | \bar b q_7 \bar q_8 q_9 | \bar b q_{10} \rangle
}
~,\nn\\
P^{\prime C}_{EW1} (1)=
{
\contraction[1ex]{\langle \bar q_1 }{q_2 }{}{\bar q_3 }
\contraction[1.9ex]{\langle }{\bar q_1 }{q_2 \bar q_3 q_4 \bar q_5 q_6 | \bar b }{q_7 }
\contraction[3ex]{\langle \bar q_1 q_2 \bar q_3 }{q_4 }{\bar q_5 q_6 | \bar b q_7 }{\bar q_8 }
\contraction[4ex]{\langle \bar q_1 q_2 \bar q_3 q_4 \bar q_5 }{q_6 }{| \bar b q_7 \bar q_8 q_9 | \bar b }{q_{10}}
\contraction[1ex]{\langle \bar q_1 q_2 \bar q_3 q_4 }{\bar q_5 }{q_6 | \bar b q_7 \bar q_8 }{q_9 }
\langle \bar q_1 q_2 \bar q_3 q_4 \bar q_5 q_6 | \bar b q_7 \bar q_8 q_9 | \bar b q_{10} \rangle
}
~&,&~~
P'_{EW1} (1)=
{
\contraction[4.4ex]{\langle \bar q_1 }{q_2 }{\bar q_3 q_4 \bar q_5 q_6 | \bar b q_7 \bar q_8 q_9 | \bar b }{q_{10}}
\contraction[5ex]{\langle }{\bar q_1 }{q_2 \bar q_3 q_4 \bar q_5 q_6 | \bar b }{q_7 }
\contraction[2.4ex]{\langle \bar q_1 q_2 \bar q_3 }{q_4 }{\bar q_5 q_6 | \bar b q_7 }{\bar q_8 }
\contraction[1ex]{\langle \bar q_1 q_2 }{\bar q_3 }{q_4 \bar q_5 }{q_6 }
\contraction[3ex]{\langle \bar q_1 q_2 \bar q_3 q_4 }{\bar q_5 }{q_6 | \bar b q_7 \bar q_8 }{q_9 }
\langle \bar q_1 q_2 \bar q_3 q_4 \bar q_5 q_6 | \bar b q_7 \bar q_8 q_9 | \bar b q_{10} \rangle
}
~,\nn\\
P^{\prime C}_{EW2} (1)=
{
\contraction[3.1ex]{\langle \bar q_1 }{q_2 }{\bar q_3 q_4 \bar q_5 q_6 | \bar b q_7 }{\bar q_8 }
\contraction[3.8ex]{\langle }{\bar q_1 }{q_2 \bar q_3 q_4 \bar q_5 q_6 | \bar b }{q_7 }
\contraction[5ex]{\langle \bar q_1 q_2 \bar q_3 }{q_4 }{\bar q_5 q_6 | \bar b q_7 \bar q_8 q_9 | \bar b }{q_{10} } \contraction[1ex]{\langle \bar q_1 q_2 }{\bar q_3 }{q_4 \bar q_5 }{q_6 }
\contraction[1.9ex]{\langle \bar q_1 q_2 \bar q_3 q_4 }{\bar q_5 }{q_6 | \bar b q_7 \bar q_8 }{q_9 }
\langle \bar q_1 q_2 \bar q_3 q_4 \bar q_5 q_6 | \bar b q_7 \bar q_8 q_9 | \bar b q_{10} \rangle
}
~&,&~~
P'_{EW2} (1)=
{
\contraction[1ex]{\langle \bar q_1 }{q_2 }{}{\bar q_3 }
\contraction[2.6ex]{\langle }{\bar q_1 }{q_2 \bar q_3 q_4 \bar q_5 q_6 | \bar b }{q_7 }
\contraction[3.9ex]{\langle \bar q_1 q_2 \bar q_3 }{q_4 }{\bar q_5 q_6 | \bar b q_7 \bar q_8 q_9 | \bar b }{q_{10}}
\contraction[1ex]{\langle \bar q_1 q_2 \bar q_3 q_4 \bar q_5 }{q_6 }{| \bar b q_7 }{\bar q_8 }
\contraction[1.6ex]{\langle \bar q_1 q_2 \bar q_3 q_4 }{\bar q_5 }{q_6 | \bar b q_7 \bar q_8 }{q_9 }
\langle \bar q_1 q_2 \bar q_3 q_4 \bar q_5 q_6 | \bar b q_7 \bar q_8 q_9 | \bar b q_{10} \rangle
}
~.\nn\\
\label{eq:ContList}
\eea
These are easy to verify from Fig.~\ref{BKpipifig}.

Recall that the momentum assignment (1) corresponds to $K(p_1)$,
$\pi_1(p_2)$ and $\pi_2(p_3)$, while (2) corresponds to $K(p_1)$,
$\pi_2(p_2)$ and $\pi_1(p_3)$.  Contractions of type (2) can be
obtained by acting with $P_{23}$, where $P_{ij}$ is the permutation
operator which exchanges the $i^{th}$ and $j^{th}$ mesons of the final
state.  For example, the contraction $C'_1(2)$ is
\beq
C'_1(2) = P_{23} C'_1(1) = 
{
\contraction[1ex]{\langle \bar q_1 q_2 \bar q_5 }{q_6 }{}{\bar q_3 }
\contraction[1ex]{\langle \bar q_1 q_2 \bar q_5 q_6 \bar q_3 }{q_4 }{| \bar b q_7 }{\bar q_8 }
\contraction[1.5ex]{\langle \bar q_1 q_2 }{\bar q_5 }{q_6 \bar q_3 q_4 | \bar b }{q_7 }
\contraction[3ex]{\langle \bar q_1 }{q_2 }{\bar q_5 q_6 \bar q_3 q_4 | \bar b q_7 \bar q_8 q_9 | \bar b }{q_{10} }
\contraction[3.5ex]{\langle }{\bar q_1 }{q_2 \bar q_5 q_6 \bar q_3 q_4 | \bar b q_7 \bar q_8 }{q_9 }
\langle \bar q_1 q_2 \bar q_5 q_6 \bar q_3 q_4 | \bar b q_7 \bar q_8 q_9 | \bar b q_{10} \rangle
} ~.
\eeq
In the same vein, contractions of type ($n$) can be obtained by acting
on contractions of type (1) with the appropriate permutation operator
(exchanges, cyclic or anti-cyclic permutations).

With these, it is straightforward to express the tree and EWP diagrams
in terms of contractions. We have
\bea
T'_j &=& \frac{|\lambda_u^{(s)}|}{\sqrt{6}} c_i \left( T'_{j,i}(1) +...+ T'_{j,i}(6) \right) ~,\nn\\
C'_j &=& \frac{|\lambda_u^{(s)}|}{\sqrt{6}} c_i \left( C'_{j,i}(1) +...+ C'_{j,i}(6) \right) ~,\nn\\
P'_{EWj} &=& -\frac{3}{2} \frac{|\lambda_t^{(s)}|}{\sqrt{6}} c_i \left( P'_{EWj,i}(1) +...+ P'_{EWj,i}(6) \right) ~,\nn\\
P^{\prime C}_{EWj} &=& -\frac{3}{2} \frac{|\lambda_t^{(s)}|}{\sqrt{6}} c_i \left( P^{\prime C}_{EWj,i}(1) +...+ P^{\prime C}_{EWj,i}(6) \right) ~,
\label{eq:diagVScontTS}
\eea
where the sum is over $i=1,2$ for trees and $i=9,10$ for EWPs.

The point is that, with a totally symmetric state, the contractions
$T'_{j,i}(m)$ and $P'_{EWj,i}(n)$ are simply different ways on writing the
same thing.  Applying the permutation operator $P_{13}$ (for example),
it is easy to show that
\bea
P_{13} T'_{j,i}(1) = P'_{EWj,i}(1)~,~~~P_{13} T'_{j,i}(2) = P'_{EWj,i}(6)~,~~~P_{13} T'_{j,i}(3) = P'_{EWj,i}(5)~,\nn\\
P_{13} T'_{j,i}(4) = P'_{EWj,i}(4)~,~~~P_{13} T'_{j,i}(5) = P'_{EWj,i}(3)~,~~~P_{13} T'_{j,i}(6) = P'_{EWj,i}(2)~.
\eea
The situation here is very similar to what we found in the $B \to K
\pi$ decay.  In that case, both $T'$ and $\pewp$ diagrams were written
in terms of $EM'$-type contractions.  Here, we use a slightly different
notation, but the above equation proves that $T'_i$ and $P'_{EWi}$
diagrams ($i=1,2$) actually contain the same type of contraction.

Similar relations exist between the $C'_{j,i}(m)$ and $P^{\prime
  C}_{EWj,i}(n)$ contractions.  Thus, assuming the Wilson coefficients
respect the approximate equality $c_1/c_2 \approx c_9/c_{10}$, it is
straightfoward to find the following SU(3) relations from
Eq.~(\ref{eq:diagVScontTS}):
\bea
P'_{EW1} = - \frac{3}{2} \frac{|\lambda_t^{(s)}|}{|\lambda_u^{(s)}|} \frac{c_9+c_{10}}{c_1+c_2} T'_1~,~~~~~~
P'_{EW2} = - \frac{3}{2} \frac{|\lambda_t^{(s)}|}{|\lambda_u^{(s)}|} \frac{c_9+c_{10}}{c_1+c_2} T'_2~,\nn\\
P^{\prime C}_{EW1} = - \frac{3}{2} \frac{|\lambda_t^{(s)}|}{|\lambda_u^{(s)}|} \frac{c_9+c_{10}}{c_1+c_2} C'_1~,~~~~~~
P^{\prime C}_{EW2} = - \frac{3}{2} \frac{|\lambda_t^{(s)}|}{|\lambda_u^{(s)}|} \frac{c_9+c_{10}}{c_1+c_2} C'_2~.
\label{exactrels}
\eea

Now, these relations assume only SU(3) symmetry and the approximate
ratio of Wilson coefficients.  The expected error due to
SU(3)-breaking effects is $O(30\%)$.  However, when all contributions
to $\btokpipi$ are taken into account, the net error is much smaller,
$O(5\%)$, since EWPs and trees are subleading effects. This is
consistent with the error estimates for EWP-tree relations in
$\btokpi$ given in Ref.~\cite{NR}.

Finally, we note that the assumption $c_1/c_2 = c_9/c_{10}$ is not
necessary.  It is actually possible to prove EWP-tree relations which
are exact under SU(3).  They are
\bea
P'_{EWi} &=& -\frac{3}{4} \frac{|\lambda_t^{(s)}|}{|\lambda_u^{(s)}|} \left[ \frac{c_9+c_{10}}{c_1+c_2} (T'_i + C'_i) + \frac{c_9-c_{10}}{c_1-c_2} (T'_i - C'_i)  \right]~,\nn\\
P^{\prime C}_{EWi} &=& -\frac{3}{4} \frac{|\lambda_t^{(s)}|}{|\lambda_u^{(s)}|} \left[ \frac{c_9+c_{10}}{c_1+c_2} (T'_i + C'_i) - \frac{c_9-c_{10}}{c_1-c_2} (T'_i - C'_i)  \right]~,
\label{exactrels2}
\eea
for $i=1,2$.  These are similar to the exact SU(3) EWP-tree relations
for $\btokpi$ given in Ref.~\cite{GPY}. (When we assume that $c_1/c_2
= c_9/c_{10}$, we recover the relations of Eq.~(\ref{exactrels}).)

\subsection{\boldmath $\pi\pi$-symmetric case}

We now consider the $\pi\pi$-symmetric state.  Applying the formalism
of contractions to $\ket{S_{\pi\pi}}$ with the effective hamiltonian
$\mathcal{H}_{eff}$, we obtain the amplitude from
\beq
\mathcal{A}(B \to K \pi_1 \pi_2)_{\pi\pi{\hbox{-}}sym} = \sum_{contractions} \bra{S_{\pi\pi}} \mathcal{H}_{eff} \ket{B}~.
\label{eq:FormCont}
\eeq
Again, there are many contractions involved.  

We use the same notation as in the previous subsection, but now the
number in parentheses only goes from 1 to 2. Thus, $X_i(1)$ ($X_i(2)$)
denotes an $X$-type contraction of operator $O_i$ of
$\mathcal{H}_{eff}$ arising from the first (second) term of the first
relation in Eq.~(\ref{eq:pipisymstate}). The expressions for the trees
and EWPs in terms of contractions are the same as for the totally
symmetric state $\ket{S}$ [Eq.~(\ref{eq:diagVScontTS})], but with only
two permutation terms:
\bea
T'_j &=& \frac{|\lambda_u^{(s)}|}{\sqrt{2}} c_i \left( T'_{j,i}(1) + T'_{j,i}(2) \right) ~,\nn\\
C'_j &=& \frac{|\lambda_u^{(s)}|}{\sqrt{2}} c_i \left( C'_{j,i}(1) + C'_{j,i}(2) \right) ~,\nn\\
P'_{EWj} &=& -\frac{3}{2} \frac{|\lambda_t^{(s)}|}{\sqrt{2}} c_i \left( P'_{EWj,i}(1) + P'_{EWj,i}(2) \right) ~,\nn\\
P^{\prime C}_{EWj} &=& -\frac{3}{2} \frac{|\lambda_t^{(s)}|}{\sqrt{2}} c_i \left( P^{\prime C}_{EWj,i}(1) + P^{\prime C}_{EWj,i}(2) \right) ~,
\label{eq:diagVScont}
\eea
where, as usual, the sum is over $i=1,2$ for trees and $i=9,10$ for EWPs.

Based on the EWP-tree relations in $\btokpi$, from the previous
equation we would expect to find a relation between $T'_j$ and
$P'_{EWj}$ (or between $C'_j$ and $P^{\prime C}_{EWj}$) under SU(3)
symmetry. And indeed, there is such a relation: for example,
$P'_{EW1}(1)$ can be obtained from $T'_1(1)$ by applying the
permutation operator $P_{13}$:
\beq
P'_{EW1}(1) = P_{13} T'_1(1) ~.
\label{eq:KpiPermu}
\eeq
The above equality can be verified easily from
Eq.~(\ref{eq:ContList}). Other pairs of contractions are related
similarly. The problem is that $P_{13}$ corresponds to the exchange of
the $K$ meson and one of the $\pi$'s. But a $K \leftrightarrow \pi$
exchange is not a valid operation here since the initial state is not
defined as being symmetric under such an exchange. More generally,
this conclusion applies to all four states of mixed symmetry.  Thus,
there are no exact SU(3) EWP-tree relations for the mixed states in $B
\to K\pi\pi$ decays. This means that, for these states, we need
different additional input in order to reduce the number of effective
diagrams. 

Fortunately, there is a possible piece of additional information. The
EWP-tree relations of Eqs.~(\ref{EWPtreerel1}) and (\ref{EWPtreerel2})
hold for $\btokpi$ to all orders in $\alpha_s$. However, it was shown
in Ref.~\cite{contractions} that one can also work order-by-order in
$\alpha_s$, i.e.\ perform the contractions analysis for processes with
0, 1, 2, etc.\ internal gluons. At leading order (LO), different
EWP-tree relations appear. As we see below, a similar behavior holds
for $\btokpipi$.

Contractions are related to Fierz tranformations ($q_7 \leftrightarrow
q_9$ in Eq.~(\ref{eq:ContList})) in the following way:
\bea
C'_{1,1} \stackrel{Fierz}{=} P'_{EW1,10}~,&&~~~~ C'_{1,2} \stackrel{Fierz}{=} P'_{EW1,9}~,\nn\\
C'_{2,1} \stackrel{Fierz}{=} P'_{EW2,10}~,&&~~~~ C'_{2,2} \stackrel{Fierz}{=} P'_{EW2,9}~,\nn\\
T'_{1,1} \stackrel{Fierz}{=} P^{\prime C}_{EW1,10}~,&&~~~~ T'_{1,2} \stackrel{Fierz}{=} P^{\prime C}_{EW1,9}~,\nn\\
T'_{2,1} \stackrel{Fierz}{=} P^{\prime C}_{EW2,10}~,&&~~~~ T'_{2,2} \stackrel{Fierz}{=} P^{\prime C}_{EW2,9}~.
\eea
That is, since Fierz relations hold at the level of operators,
contractions of operators $O_{1,2}$ are related to those of operators
$O_{10,9}$ respectively .  Applying this to diagram $P'_{EW1}$ of
Eq. (\ref{eq:diagVScont}) for example, we have
\bea
P'_{EW1} &=& -\frac{3}{2} \frac{|\lambda_t^{(s)}|}{\sqrt{2}} c_i \left( P'_{EW1,i}(1) + P'_{EW1,i}(2) \right)\nn\\
&=& -\frac{3}{2} \frac{|\lambda_t^{(s)}|}{\sqrt{2}} \left( c_9 P'_{EW1,9}(1) + c_{10} P'_{EW1,10}(1) + c_9 P'_{EW1,9}(2) + c_{10} P'_{EW1,10}(2) \right) \nn\\
&\stackrel{Fierz}{=}& -\frac{3}{2} \frac{|\lambda_t^{(s)}|}{\sqrt{2}} \left( c_9 C'_{1,2}(1) + c_{10} C'_{1,1}(1) + c_9 C'_{1,2}(2) + c_{10} C'_{1,1}(2) \right)~.
\label{eq:crossed1}
\eea
We also have
\bea
C'_1 &=& \frac{|\lambda_u^{(s)}|}{\sqrt{2}} c_i \left( C'_{1,i}(1) + C'_{1,i}(2) \right)\nn\\
&=& \frac{|\lambda_u^{(s)}|}{\sqrt{2}} \left( c_1 C'_{1,1}(1) + c_2 C'_{1,2}(1) + c_1 C'_{1,1}(2) + c_2 C'_{1,2}(2) \right) ~.
\eea
We therefore see that $P'_{EW1}$ is not proportional to the $C'_{1}$.
It would be if the Wilson coefficients respected the equality $c_1/c_2
= c_{10}/c_9$, but this obviously does not hold (what is true is that
$c_1/c_2 \approx c_9/c_{10}$).

This can be ameliorated by working only to LO in $\alpha_s$.  In this
case, color effects can be extracted out \cite{contractions}, so that
the above equations become
\bea
P'_{EW1} &=& -\frac{3}{2} \frac{|\lambda_t^{(s)}|}{\sqrt{2}} \left( c_9 N_c^2 \overline{C'_{1}}(1) + c_{10} N_c \overline{C'_{1}}(1) + c_9 N_c^2 \overline{C'_{1}}(2) + c_{10} N_c \overline{C'_{1}}(2) \right)\nn\\
&=& -\frac{3}{2} \frac{|\lambda_t^{(s)}|}{\sqrt{2}} \left( c_9 N_c^2 + c_{10} N_c \right) \left( \overline{C'_{1}}(1) + \overline{C'_{1}}(2)\right) + O(\alpha_s) ~, \nn\\
C'_1 &=& \frac{|\lambda_u^{(s)}|}{\sqrt{2}} \left( c_1 N_c \overline{C'_{1}}(1) + c_2 N_c^2 \overline{C'_{1}}(1) + c_1 N_c \overline{C'_{1}}(2) + c_2 N_c^2 \overline{C'_{1	}}(2) \right)\nn\\
&=& \frac{|\lambda_u^{(s)}|}{\sqrt{2}} \left( c_1 N_c + c_2 N_c^2 \right) \left( \overline{C'_{1}}(1) + \overline{C'_{1}}(2) \right) + O(\alpha_s) ~, 
\label{eq:crossed2}
\eea
in which $N_c=3$ is the number of colors in QCD and the overline
notation indicates color-extracted contractions. We therefore obtain
the relation
\beq
P'_{EW1} \approx -\frac{3}{2} \frac{|\lambda_t^{(s)}|}{|\lambda_u^{(s)}|} \frac{c_9 + c_{10}/N_c}{c_1/N_c + c_2} \, C'_1 ~,
\eeq
which is valid at LO and under isospin symmetry (SU(3) was not used
above).  The same procedure can be applied to other diagrams, with the
result that
\bea
P'_{EW2} &\approx & -\frac{3}{2} \frac{|\lambda_t^{(s)}|}{|\lambda_u^{(s)}|} \frac{c_9 + c_{10}/N_c}{c_1/N_c + c_2} \,C'_2~,\nn\\
P^{\prime C}_{EW1} &\approx & -\frac{3}{2} \frac{|\lambda_t^{(s)}|}{|\lambda_u^{(s)}|} \frac{c_9/N_c + c_{10}}{c_1 + c_2/N_c} \,T'_1~,\nn\\
P^{\prime C}_{EW2} &\approx & -\frac{3}{2} \frac{|\lambda_t^{(s)}|}{|\lambda_u^{(s)}|} \frac{c_9/N_c + c_{10}}{c_1 + c_2/N_c} \,T'_2~.
\label{crossedrels}
\eea
We refer to these as ``crossed'' EWP-tree relations.

As noted above, the crossed relations hold only at LO -- these are not
reproduced by the higher-order diagrams. The error is therefore
$O(\alpha_s)$. The size of this error then depends crucially on what
the value of $\alpha_s$ is for this calculation. For example, if soft
gluons are important, then $\alpha_s$ is large, and the use of these
relations is not a good approximation. To address this question, we
rely on theoretical input. There are basically three approaches used
in calculations of hadronic $B$ decays: QCD factorization (QCDf)
\cite{QCDf}, perturbative QCD (pQCD) \cite{pQCD}, and soft collinear
effective theory (SCET) \cite{SCET}.  All three methods perform their
studies of two-body decays by taking the $m_b \to \infty $ limit and
separating the nonperturbative low-energy (soft) effects from those at
high energies (hard effects)\footnote{In fact, there are three energy
  scales for gluons: $\Lambda_{QCD}$ (soft), $m_b$ (hard), and
  $\sqrt{\Lambda_{QCD} \, m_b}$ (hard-collinear). The presence of
  these three scales affects calculations within a specific model, but
  does not change our conclusions regarding the value of $\alpha_s$ in
  the expansion.}.  All gluons (soft and hard) between two quarks in
the same meson are absorbed into the parameters describing
hadronization (decay constants).  Other gluons between quarks of two
different mesons are absorbed into the form factors.  For the
remaining gluons, in all three approaches it was found that soft
gluons are suppressed, so that $\alpha_s = \alpha_s(m_b) \sim 20\%$
\cite{contractions}. This permits an expansion in $\alpha_s$, and this
was done in QCDf, pQCD, and SCET.  Here we assume that this also holds
in the case of three-body decays, so that the use of crossed EWP-tree
relations is, in fact, a reasonable approximation. (There have been
some studies of such decays, and they support this assumption
\cite{3bodystudies}.)

The relations suffer from an additional error due to the fact that the
ratio of Wilson coefficients is strongly dependent on the choice of
renormalization scale.  This effect can be taken into account by
considering a large range of values for this ratio. On the whole, we
estimate that the total error is roughly comparable to that when SU(3)
is assumed, $O(30\%)$. However, since EWPs and trees are themselves
subleading effects in $\btokpipi$ decays, the net effect in concrete
applications is much smaller, $O(5\%)$.

\section{Measuring $\gamma$}

The EWP-tree relations found above do indeed permit the extraction of
$\gamma$ from $\btokpipi$ decays. However, the precise procedure used
depends on what the $K\pi\pi$ state is. One can use only events
corresponding to the totally symmetric final state $\ket{S}$. Or one
can combine the two $S_3$ states, both symmetric under the exchange of
the two pions, whose sum forms $\ket{S_{\pi\pi}}$
[Eq.~(\ref{eq:pipisymstate})].

If the $K\pi\pi$ state is $\ket{S}$, the exact SU(3) EWP-tree
relations [Eq.~(\ref{exactrels})] hold. For the effective diagrams,
this implies that
\beq
P'_{EW,b} = - \frac{3}{2} \frac{|\lambda_t^{(s)}|}{|\lambda_u^{(s)}|} \frac{c_9+c_{10}}{c_1+c_2} \, T'_b~.
\eeq
Thus, there are only five effective diagrams in the six $\btokpipi$
decay amplitudes. This corresponds to 10 theoretical parameters: 5
magnitudes of diagrams, 4 relative (strong) phases, and $\gamma$.
Since there are 11 experimental observables, $\gamma$ can be extracted
by doing a fit.

The fit itself is somewhat unusual. All experimental observables are
momentum dependent, as are the diagrams. In obtaining the best-fit
``values'' of the diagrams, one will determine the momentum dependence
of their magnitudes and relative strong phases. However, $\gamma$ is
independent of the particles' momenta. Thus, the fit must yield a
momentum-independent value for $\gamma$. The error on $\gamma$ must
take into account any momentum dependence.

Now, the extraction of the decay amplitudes from the Dalitz plots is
rather difficult, and has a certain amount of input -- distributions
of resonant effects (e.g.\ Breit-Wigner), treatment of non-resonant
contributions, etc. It is possible the input chosen is imprecise, and
can lead to a momentum-dependent value for $\gamma$. In this sense,
the requirement that $\gamma$ be momentum independent may provide some
hints regarding the form of the decay amplitudes.

In the $\pi\pi$-symmetric case, we have shown above that there are no
exact SU(3) EWP-tree relations for $\ket{S_{\pi\pi}}$. However, the
crossed EWP-tree relations [Eq.~(\ref{crossedrels})] do hold. By
applying these to the effective diagrams of Eq.~(\ref{eq:effdiag}) we
have
\bea
P'_{EW,a} \approx - \frac{3}{2} \frac{|\lambda_t^{(s)}|}{|\lambda_u^{(s)}|} \frac{c_9/N_c + c_{10}}{c_1 + c_2/N_c} \, T'_a~.
\eea
Once again, the number of effective diagrams is reduced to five, which
makes the extraction of $\gamma$ possible. One can even use both the
exact and crossed EWP-tree relations, in which case fewer observables
are needed to obtain $\gamma$.

In both cases, the theoretical error is $O(5\%)$. The advantage of
using $\ket{S_{\pi\pi}}$ rather than $\ket{S}$ is that the number of
events is somewhat larger.

\section{Conclusions}

It has been known for some time that there are relations between the
electroweak-penguin (EWP) and tree contributions to $\btokpi$ decays.
In particular, apart from the weak phases, the diagrams $\pewp$ and
$\pewcp$ are proportional to $T'$ and $C'$, respectively, to a good
approximation. In 2003, Deshpande, Sinha and Sinha (DSS) attempted to
use these EWP-tree relations to extract $\gamma$ from $\btokpipi$
decays. Working with the $\pi\pi$-symmetric $K\pi\pi$ states
($\ket{S_{\pi\pi}}$), they noted that $B^+ \to K^0\pi^+\pi^0$ receives
only tree and EWP contributions. DSS' assumption was that these are
related as in $\btokpi$, and this additional information allowed
$\gamma$ to be obtained. Unfortunately, it was subsequently shown that
the EWP-tree relation does not hold, so that $\gamma$ cannot be
extracted using DSS' method.

In this paper, we revisit the question of measuring $\gamma$ in
$\btokpipi$ decays, and we show that, in fact, it is possible. We
first define the diagrams contributing to $\btokpipi$. Because there
are three particles in the final state, there are two types of each
diagram. We call them $T'_1$, $T'_2$, $P'_{EW1}$, $P'_{EW2}$, etc. We
then express each $\btokpipi$ amplitude in terms of these diagrams.
DSS' assumption is that $P'_{EW2} + P^{\prime C}_{EW1}$ is
proportional to $T'_1 + C'_2$. Using the contractions formalism, we
are able to express all diagrams in terms of contractions, and thereby
show that there are, in fact, EWP-tree relations. To be specific, we
find that $P'_{EWi} \propto T'_i$ ($i=1,2$) and $P^{\prime C}_{EWi}
\propto C'_i$ ($i=1,2$). From this, we see immediately that the DSS
assumption is indeed false.

Now, when one writes the amplitudes in terms of diagrams, one sees
that there are more unknown theoretical parameters than there are
observables, so that weak-phase information cannot be obtained without
additional input. The EWP-tree relations provide this input, and allow
$\gamma$ to be measured in $\btokpipi$ decays. But there is a
complication. EWP-tree relations require flavor SU(3) symmetry. Since
$K$ and $\pi$ are equivalent under this symmetry, one has to deal with
three identical particles in the $K\pi\pi$ final states. The
permutation group of three objects is $S_3$, which has as eigenstates
a totally symmetric state of the three objects, a totally
antisymmetric state, and four mixed states.  However, since the
relative angular momentum between the particles is not fixed (due to
the fact that we have a three-particle final state), the state is not
necessarily symmetric under permutations of the mesons. On the other
hand, it turns out that the EWP-tree relations apply only to the
totally symmetric state ($\ket{S}$). Thus, if one wants to apply these
relations, one must isolate those events which correspond to the state
$\ket{S}$.

The state $\ket{S_{\pi\pi}}$ is a combination of $\ket{S}$ and one of
the $S_3$ mixed states. As such, the above EWP-tree relations do not
apply to it. Fortunately, there is an alternative. If one works to
leading order (LO) in $\alpha_s$, we find ``crossed'' EWP-tree
relations: $P^{\prime C}_{EWi} \propto T'_i$ ($i=1,2$) and $P'_{EWi}
\propto C'_i$ ($i=1,2$). We expect these to hold approximately, since
$\alpha_s(m_b) \sim 0.2$. The crossed EWP-tree relations can be used
with $\ket{S_{\pi\pi}}$. They are valid under isospin symmetry --
SU(3) is not used.

In both cases, we estimate the theoretical error to be $O(5\%)$.
Experimentally, one can choose to use either state. The advantage of
$\ket{S}$ over $\ket{S_{\pi\pi}}$ is that the EWP-tree relations are
exact, as opposed to LO. On the other hand, the advantage of
$\ket{S_{\pi\pi}}$ over $\ket{S}$ is that the number of events is
somewhat larger.

\bigskip
\noindent
{\bf Acknowledgments}:
We thank M. Gronau, J. Rosner, R. Sinha, R. MacKenzie and A. Soffer
for helpful communications, and A. Datta for collaboration in the
beginning stages of this project. This work was financially supported
by NSERC of Canada and FQRNT of Qu\'ebec.




\begin{thebibliography}{99}

\bibitem{pdg}
  C.~Amsler {\it et al.}  [Particle Data Group],
  Phys.\ Lett.\  B {\bf 667}, 1 (2008).

\bibitem{NQ} Y.~Nir and H.~R.~Quinn,
  Phys.\ Rev.\ Lett.\  {\bf 67}, 541 (1991).

\bibitem{DH} N.~G.~Deshpande and X.~G.~He,
  Phys.\ Rev.\ Lett.\  {\bf 74}, 26 (1995)
  [Erratum-ibid.\  {\bf 74}, 4099 (1995)]
  [arXiv:hep-ph/9408404].

\bibitem{NR} M.~Neubert and J.~L.~Rosner,
  Phys.\ Lett.\  B {\bf 441}, 403 (1998)
  [arXiv:hep-ph/9808493],
  Phys.\ Lett.\  B {\bf 441}, 403 (1998)
  [arXiv:hep-ph/9808493].

\bibitem{GPY} M.~Gronau, D.~Pirjol and T.~M.~Yan,
  Phys.\ Rev.\  D {\bf 60}, 034021 (1999)
  [Erratum-ibid.\  D {\bf 69}, 119901 (2004)]
  [arXiv:hep-ph/9810482].

\bibitem{Kpisol} M.~Imbeault, A.~L.~Lemerle, V.~Page and D.~London,
  Phys.\ Rev.\ Lett.\  {\bf 92}, 081801 (2004)
  [arXiv:hep-ph/0309061].

\bibitem{LNQS} H.~J.~Lipkin, Y.~Nir, H.~R.~Quinn and A.~Snyder,
  Phys.\ Rev.\  D {\bf 44}, 1454 (1991).

\bibitem{DSS} N.~G.~Deshpande, N.~Sinha and R.~Sinha,
  Phys.\ Rev.\ Lett.\  {\bf 90}, 061802 (2003)
  [arXiv:hep-ph/0207257].

\bibitem{Grocomment} M.~Gronau,
  Phys.\ Rev.\ Lett.\  {\bf 91}, 139101 (2003)
  [arXiv:hep-ph/0305144].

\bibitem{BPPP} N.~Rey-Le Lorier, M.~Imbeault and D.~London,
  arXiv:1011.4972 [hep-ph].

\bibitem{GHLR} M.~Gronau, O.~F.~Hernandez, D.~London and J.~L.~Rosner,
  Phys.\ Rev.\ D {\bf 50}, 4529 (1994), Phys.\ Rev.\ D {\bf 52}, 6374
  (1995).

\bibitem{BBL}
  G.~Buchalla, A.~J.~Buras and M.~E.~Lautenbacher,
  Rev.\ Mod.\ Phys.\  {\bf 68}, 1125 (1996)
  [arXiv:hep-ph/9512380].

\bibitem{contractions}
  M.~Imbeault, A.~Datta and D.~London,
  Int.\ J.\ Mod.\ Phys.\  A {\bf 22}, 2057 (2007)
  [arXiv:hep-ph/0603214].

\bibitem{KpipiSU3} However, for a discussion of $\btokpipi$ decays
  with SU(3) symmetry, see
M.~Ciuchini, M.~Pierini and L.~Silvestrini,
  Phys.\ Rev.\  D {\bf 74}, 051301 (2006)
  [arXiv:hep-ph/0601233],
  Phys.\ Lett.\  B {\bf 645}, 201 (2007)
  [arXiv:hep-ph/0602207];
M.~Gronau, D.~Pirjol, A.~Soni and J.~Zupan,
  Phys.\ Rev.\  D {\bf 75}, 014002 (2007)
  [arXiv:hep-ph/0608243].

\bibitem{BS}
  A.~J.~Buras and L.~Silvestrini,
  Nucl.\ Phys.\  B {\bf 569}, 3 (2000)
  [arXiv:hep-ph/9812392].

\bibitem{GR2005} M.~Gronau and J.~L.~Rosner,
  Phys.\ Rev.\  D {\bf 72}, 094031 (2005)
  [arXiv:hep-ph/0509155].

\bibitem{QCDf} M.~Beneke, G.~Buchalla, M.~Neubert and C.~T.~Sachrajda,
  Phys.\ Rev.\ Lett.\  {\bf 83}, 1914 (1999)
  [arXiv:hep-ph/9905312],
  Nucl.\ Phys.\  B {\bf 591}, 313 (2000)
  [arXiv:hep-ph/0006124],
  Nucl.\ Phys.\  B {\bf 606}, 245 (2001)
  [arXiv:hep-ph/0104110].

\bibitem{pQCD} Y.~Y.~Keum, H.~n.~Li and A.~I.~Sanda,
  Phys.\ Lett.\  B {\bf 504}, 6 (2001)
  [arXiv:hep-ph/0004004],
  Phys.\ Rev.\  D {\bf 63}, 054008 (2001)
  [arXiv:hep-ph/0004173],

\bibitem{SCET} C.~W.~Bauer, S.~Fleming, D.~Pirjol and I.~W.~Stewart,
  Phys.\ Rev.\  D {\bf 63}, 114020 (2001)
  [arXiv:hep-ph/0011336].

\bibitem{3bodystudies} C.~H.~Chen and H.~n.~Li,
  Phys.\ Lett.\  B {\bf 561}, 258 (2003)
  [arXiv:hep-ph/0209043];
C.~H.~Chen,
  Phys.\ Lett.\  B {\bf 560}, 178 (2003)
  [arXiv:hep-ph/0301154].

\end{thebibliography}
\end{document}